\documentclass[aps,amsmath,amssymb,amsfonts, superscriptaddress,twocolumn,longbibliography]{revtex4-1}
\usepackage{graphicx}
\usepackage{graphics}
\usepackage{dcolumn}
\usepackage[colorlinks, linkcolor= blue, citecolor = blue, urlcolor=blue]{hyperref}
\usepackage{multirow}
\usepackage{bm}
\usepackage{latexsym}
\usepackage{amssymb}
\usepackage{amsmath}
\usepackage{amsfonts}
\usepackage{layout}
\usepackage{verbatim}
\usepackage{epsfig}
\usepackage{graphicx}
\usepackage{amsbsy}
\usepackage{xcolor}
\usepackage[caption=false]{subfig}
\newcommand{\bea}{\begin{eqnarray*}}
	\newcommand{\eea}{\end{eqnarray*}}
\newcommand{\bne}{\begin{equation*}}
\newcommand{\ede}{\end{equation*}}

\newcommand{\bnen}{\begin{equation}}
\newcommand{\eden}{\end{equation}}
\newcommand{\bean}{\begin{eqnarray}}
\newcommand{\eean}{\end{eqnarray}}
\newcommand{\bsen}{\begin{subequations}}
	\newcommand{\esen}{\end{subequations}}

\newcommand{\bna}{\begin{array}}
	\newcommand{\eda}{\end{array}}
\newcommand{\bnm}{\begin{enumerate}}
	\newcommand{\edm}{\end{enumerate}}

\begin{document}
 
\title{Orbital angular momentum of Bloch electrons: equilibrium formulation, magneto-electric phenomena, and the orbital Hall effect}

\author{Rhonald Burgos Atencia}
\email{rhburgos@gmail.com}
\affiliation{School of Physics, The University of New South Wales, Sydney 2052, Australia}
\affiliation{ARC Centre of Excellence in Low-Energy Electronics Technologies, UNSW Node, The University of New South Wales, Sydney 2052, Australia}
\author{Amit Agarwal}
\email{amitag@iitk.ac.in}
\affiliation{Department of Physics, Indian Institute of Technology, Kanpur-208016, India}
\author{Dimitrie Culcer}
\email{d.culcer@unsw.edu.au}
\affiliation{School of Physics, The University of New South Wales, Sydney 2052, Australia}
\affiliation{ARC Centre of Excellence in Low-Energy Electronics Technologies, UNSW Node, The University of New South Wales, Sydney 2052, Australia}

\begin{abstract}
The investigation of orbital angular momentum (OAM) of delocalised Bloch electrons has advanced our understanding of magnetic, transport, and optical phenomena in crystals, drawing widespread interest across various materials science domains, from metals and semiconductors to topological and magnetic materials. Here, we review OAM dynamics in depth, focusing on key concepts and non-equilibrium systems, and laying the groundwork for the thriving field of {\it orbitronics}. We review briefly the conventional understanding of the equilibrium OAM based on the modern theory of orbital magnetisation. Following this, we explore recent theoretical and experimental developments in out-of-equilibrium systems. We focus on the generation of an OAM density via the orbital magneto-electric, or Edelstein effect, the generation of an OAM current via the orbital Hall effect, the orbital torque resulting from them, along with their reciprocal non-equilibrium counterparts -- the inverse orbital Edelstein and inverse orbital Hall effects, as well as OAM conservation. We discuss the most salient achievements and the most pressing challenges in this rapidly evolving field, and in closing we highlight the future prospects of {\it orbitronics}. 
\end{abstract}
\date{\today}
\maketitle

\section{Introduction} 

Magnetic interactions involve the spin and orbital angular momentum of localised and itinerant fermions. The spin magnetic moment is well understood. In contrast, the orbital magnetic moment is well understood in localised systems, where it is envisaged as current loops formed by electrons in atomic orbits. Under certain circumstances, the crystal field of the lattice distorts the current loops and quenches the orbital magnetic moment, leaving the spin magnetic moment as the only surviving contribution. This picture forms the starting point for our understanding of magnetism in insulators and many itinerant ferromagnets.

The orbital magnetic moment of itinerant Bloch electrons, associated with their orbital angular momentum (OAM), is considerably less well understood than the orbital magnetic moment of localised systems or the spin magnetic moment. The earliest references we are aware of describing the equilibrium OAM of Bloch electrons are due to Yafet in the contexts of hyperfine interaction \cite{Yafet1961} and spin relaxation \cite{Yafet-1963}. The topic became an active research area in the 1990s when it was shown that the itinerant OAM is in part related to the Berry curvature \cite{ChangMing1995, ChangMingChe1996, GaneshSundaram1999, MingCheChang2008, Fuchs2010}, ultimately leading to the modern theory of orbital magnetisation and orbital polarisation \cite{Mokrousov_ModernMag_PRB2016, Vanderbilt2018}.

The modern theory represents the current consensus on the equilibrium properties of the OAM. The theory may be formulated in the language of extended Bloch wave functions or equivalently in that of Wannier functions. The latter formulation has gained significant popularity since the development of maximally localized Wannier functions \cite{MarzariNicola2012}. Different insights into the nature of the OAM emerge from the Wannier and Bloch formulations of the modern theory. The Wannier picture reveals two distinct mechanisms contributing to the OAM. The first mechanism, which we shall denote succinctly as the \textit{atomic} OAM, represents an OAM inherited by the itinerant electrons from the parent atomic orbitals. It has been referred to as the intra-atomic contribution \cite{PezoArmando2022, PezoArmando2023} and can be regarded as an infinite mass approximation. However, even if this atomic contribution is absent, an \textit{itinerant} OAM can still exist -- this has been referred to recently as the inter-atomic contribution \cite{BuschOliver2023, PezoArmando2023}, and can be seen as a finite-mass correction. It is understood at present that both the atomic and itinerant OAM need to be taken into account -- they have been considered on the same footing in recent studies \cite{CysneTarik2022, BuschOliver2023, PezoArmando2023}. In contrast, the Bloch picture of the OAM is best visualised by considering the motion of a delocalised wave packet through the lattice: part of the OAM stems from the rotation of the wave packet around its centre of mass, while another part arises from the semiclassical correction to the density of states in a magnetic field \cite{XiaoDi2005, XiaoDi2010}. Interestingly, whereas the final expressions for the equilibrium OAM in the Wannier and Bloch pictures are in agreement, individual contributions can usually not be placed in correspondence with each other. At the same time the study of OAM in centrosymmetric systems has occasionally relied on analytical ${\bm k} \cdot {\bm p}$ frameworks based on an effective OAM originating in the atomic degrees of freedom \cite{Bernevig2005, SeungynHan2022, ChoiYoungGwan2023}. At present it is not clear how this effective OAM is related to the OAM in the modern theory.

The primary focus of this review is on the \textit{non-equilibrium} properties of the OAM, which reflects the extraordinary progress in the study of OAM transport and generation by an electric field on both the theoretical and experimental fronts. The term \textit{orbitronics} encompasses the non-equilibrium dynamics of the OAM and its applications to electrical and spin-based devices \cite{DongwookGo2021}. Even though a net OAM in equilibrium requires the system to break time reversal, transport and generation of OAM are often investigated in time-reversal invariant systems. The field of orbitronics began with the seminal proposal for atomic-OAM transport by valence band holes in silicon \cite{Bernevig2005}. As the OAM current flows perpendicular to the applied electric field, this phenomenon was termed the orbital Hall effect (OHE). Following the proposal in Si, the OHE due to atomic-OAM has been investigated in many materials, primarily metals, where it is exceedingly strong \cite{Tanaka2008, Kontani2008, Kontani2009, GoDongwook2018}. It was later pointed out that the itinerant-OAM can similarly lead to an OHE. It may also be the primary mechanism responsible for the valley-Hall effect observed in graphene and transition metal dichalcogenides \cite{BhowalSayantika2021}. The OHE does not require inversion or time-reversal symmetry breaking, in the same manner as the spin-Hall effect. It was subsequently realised that an electric field can also generate a net OAM density. In analogy with spin dynamics in semiconductors ~\cite{EDELSTEIN1990}, this phenomenon has been named the orbital Edelstein effect or the orbital magneto-electric effect. The names have been used loosely, although the term \textit{Edelstein effect} tends to involve dissipation, present through the scattering time $\tau$. For the sake of conciseness in this review, we shall adhere to the latter convention and refer to OAM generation by an electric field as the orbital magneto-electric effect, or OME, regardless of whether it involves dissipation. Unlike the OHE, the OME is subject to symmetry restrictions and occurs in gyrotropic systems, in the same manner as its spin counterpart~\cite{EDELSTEIN1990}. It will not occur in centro-symmetric systems. We summarise basic discrete symmetry requirements for the OME and OHE in 
table~\eqref{tab:symforOHEandOHE} and defer specific discussion to later sections.

In light of the above, most of the research on atomic-OAM has focussed on the OHE, where large signals have been observed recently \cite{LeeDongjoon2021, DuttaSutapa2022, ChoiYoungGwan2023, GiacomoSala2023, HayashiHiroki2023, IgorLyalin2023}. Beyond giving rise to the OHE and OME, the OAM affects semi-classical quantization \cite{MingCheChang2008, GaneshSundaram1999, Fuchs2010}, contributes to magnetisation in certain time-reversal breaking materials \cite{HankeJanPhillipp2017, KimJunyeon2021, Lahiri2022}, and affects the Zeeman splitting of Dirac materials 
\cite{Rostami2015, GuitierrezRubio2016, EichMarius2018, Overweg2018, AngelikaKnothe2018, Kursmann-Ensslin-2019, YongjinLee2020, Tong2021}. It plays an essential role in linear \cite{FaridiAzadeh2023, Jianhui2024} and non-linear magneto-resistance \cite{Lahiri2022}, linear \cite{BhowalSayantika2021} and non-linear valley-Hall effects \cite{KamalDas2023}, the anomalous Nernst effect \cite{XiaoDi2006}, the gyrotropic magneto-electric effect \cite{Tsirkin2018, WangYanQi2020}, and electrical operation of orbital degrees of freedom in systems without spin-orbit coupling \cite{JoDaegeun2018, Zheng2020, XueFei2020, JohanssonAnnika2021, LeeSoogil2021, KobayashiShingo2021, ShaoQiming2021, SeifertTom2023, ChoiYoungGwan2023, WangPing2023}. In addition, the gyrotropic magnetic effect -- the generation of a current density due to a varying magnetic field -- can be expressed in terms of the intrinsic OAM and can be related to an OAM generated by an electric field~\cite{ZhongShudan2016}.

\begin{table}[tbp]
\begin{ruledtabular}
\begin{tabular}{ l c c c c }
\textrm{Effect} & $\mathcal{P}$ & $\mathcal{T}$ & $\mathcal{P}\mathcal{T}$ & \\
\colrule
OME $\sim \tau $      &    x & $\checkmark$ & x \\
OME $\sim \tau^{{0}}$ &    x & x &  $\checkmark$  \\
OHE $\sim \tau^{{0}}, \tau$ & $\checkmark$ & $\checkmark$ & $\checkmark$ \\
\end{tabular}
\end{ruledtabular}
\caption{ Relation of the OME and OHE to time reversal $\mathcal{T}$ and 
inversion $\mathcal{P}$ symmetries.\label{tab:symforOHEandOHE}
}
\end{table}

The electrical manipulation of magnetic degrees of freedom is the primary technological motivation behind the recent interest in the OAM. Most of the current interest in the OAM traces its origins to spintronics, \cite{DasSarma2004, FertAlbert2008} which focuses on manipulating the magnetisation ($\bm M$) of a ferromagnet by electrical means. A ferromagnet is typically placed next to a material that can transfer spin angular momentum ($\bm S $) to it, which involves the exchange interaction between the local magnetisation in the ferromagnet and the non-equilibrium electron spin density in the adjacent material ($H_{ex}\sim \bm S\cdot \bm M$). Generating a non-equilibrium spin density and transferring it to a ferromagnet -- charge-to-spin conversion -- is accomplished by the spin Hall or Edelstein effects, both requiring strong spin-orbit coupling. Consequently, it was primarily heavy metals such as Pt, W, Ta that were believed to be useful for spintronics \cite{DasSarma2004, Castel2012, LuqiaoLiu2012, SinovaJairo2015, ZhangWei2013, Stamm2017, YuR2018, Sagasta2018}.
Light elements, having weak spin-orbit coupling, generally took a back seat in spintronics. However, the possibility of using light elements for applications would be extremely beneficial from the perspective of mass-producing devices. Interestingly, recent experiments on Titanium (Ti) \cite{ChoiYoungGwan2023}, Chromium (Cr) \cite{IgorLyalin2023}, and light element bilayers adjacent to an FM \cite{HayashiHiroki2023} reveal a torque on $\bm M$, and suggest a magnetic mechanism beyond spin, likely the OAM. The OHE injects angular momentum into a ferromagnet and exerts a torque on its magnetisation, referred to as the orbital torque (OT) \cite{ChenXi2018, GoDongwook2020-II, DingShilei2020, Zheng2020, LeeSoogil2021, LeeDongjoon2021, KimJunyeon2021, SameerGrover2022}. At the same time, the OAM cannot exert a torque directly on $\bm M$ since there is no exchange interaction between the two. One surmises that spin-orbit coupling in the vicinity of the interface converts the OAM current into a spin current, which then exerts a torque on $\bm M$. The process implies a \textit{charge-to-orbital-to-spin} conversion before the torque is exerted. The microscopic mechanisms leading to charge-to-orbital-to-spin conversion in the vicinity of interfaces are far from being understood. The working hypothesis at present is that a non-equilibrium OAM density is first induced via OME or OHE, then it is converted to spin magnetisation, before it finally acts on the magnetisation of an adjacent FM. However, the details of this process are one of the great unanswered questions in the field.

In addition, the study of non-equilibrium OAM is in its inception, with existing studies overwhelmingly based on the equilibrium picture, that is, the non-equilibrium values of OAM-related observables are typically determined by considering the equilibrium OAM and non-equilibrium changes in occupation. The dynamics of OAM in electric and magnetic fields are not understood, 
conservation is not understood -- for example it is not clear whether the atomic and itinerant OAMs are separately conserved -- likewise, the effects of disorder and inhomogeneities are not understood. These challenges, coupled with the impressive experimental progress registered at present, will motivate research in this field for decades to come. 

We have organized this review as follows. 
In Sec.~\eqref{Sec:equilibriumOAM} we describe the theory of the equilibrium OAM, focusing on the atomic-OAM in centro-symmetric systems and itinerant-OAM in non-centrosymmetric systems. For the latter, we outline the semi-classical Bloch and Wannier pictures. We devote Sec.~\eqref{Sec:nonequilibriumeffects} to the theory of non-equilibrium OAM effects. Sec.~\eqref{Sec:OEE} covers the orbital magneto-electric effect, while Sec.~\eqref{Sec:OHE} discusses the orbital Hall effect. In Sec.~\eqref{Sec:conservationOAM}, we address the conservation of OAM. In Sec.~\eqref{Sec:Experiments}, we survey the experimental observation of the orbital magneto-electric and Hall effects as well as their reciprocal counterparts, namely the inverse orbital magneto-electric effect (IOME) and the inverse orbital Hall effect (IOHE). In Sec. \eqref{Sec:summary}, we conclude with the main fundamental challenges in the field and an outlook on future directions.

\section{Theory of equilibrium OAM}
\label{Sec:equilibriumOAM}

In this section, we will concentrate on the theoretical understanding of the equilibrium OAM, which took half a century to reach maturity and is now known as the \textit{modern theory} of orbital magnetization. The modern theory in principle provides a complete description of the equilibrium OAM in a clean system. It can be discussed in the framework of Wannier functions, which can be localised around atomic sites, or that of delocalised Bloch functions. Within the modern theory one typically distinguishes two contributions to the equilibrium OAM. In the language of Wannier functions these can be understood as an atomic contribution, associated with the OAM inherent in the atomic orbitals, and an itinerant contribution due to hopping between atoms \cite{Thonhauser2005}. In the language of Bloch wave functions these contributions are interpreted respectively as the motion of a wave packet around its centre of mass and a Berry curvature correction to the density of states, or spectral weight transfer due to the Berry curvature. At the same time, an effective OAM description underlies analytical theories designed for centro-symmetric systems. The exact relationship between these effective descriptions will need to be clarified in the future.

\subsection{Modern theory of OAM}

Theories of the itinerant-OAM face the problem of dealing with the position operator present in the definition of the OAM: the expectation value of the position operator is not well defined in extended systems with periodic boundary conditions. The semiclassical method and the quantum mechanical method based on Bloch functions tackle this problem in similar ways and are closely related since the semiclassical method relies on wave packets of Bloch functions. They both make use of matrix elements of the position operator in momentum space \cite{CallawayBook}. In the semiclassical method, the Peierls substitution mixes position and momentum, and the gradient expansion of the Hamiltonian yields a Zeeman energy correction, which has the form of a magnetic moment. In the quantum mechanical method, one defines the operator ${\bm L} = 1/2 ({\bm r} \times {\bm v} - {\bm v} \times {\bm r})$, projects it onto Bloch functions, and obtains the equilibrium OAM for a Bloch band. This has the same form as the magnetic moment found using the semiclassical theory, except for the prefactor $-e$. Strictly speaking ${\bm L}$ as defined above refers to the orbital moment, rather than the OAM, which contains an additional factor of the mass. Nevertheless it is more convenient to work with ${\bm L}$ as we have defined it given the broad interest in graphene and topological materials, where the mass is not a good parameter, and then make material specific adjustments to recover the magnetic properties.

\subsubsection{Semiclassical and Bloch pictures}

In the semiclassical picture, the Hamiltonian is expanded about the center of the wave packet $\bm r_c$:
\begin{equation}
\label{Eq:semiclassicalapproach}
H=H_c + \frac{1}{2}\left[(\bm r - \bm r_c) \cdot {\bm \nabla} H + {\bm \nabla} H \cdot (\bm r - \bm r_c) \right].   
\end{equation}
The local Hamiltonian $H_c$ satisfies the eigenvalue equation 
$ H_c | \psi_{\bm q}(\bm r_c) \rangle = \varepsilon_c (\bm q) | \psi_{\bm q}(\bm r_c) \rangle$, where $\bm q$ is the Bloch wave vector. 
The eigenfunctions of the local Hamiltonian are used as the basis to expand the wave packet $| \Psi \rangle $. 
The expectation value of the local Hamiltonian $H_c$ gives the band energy $\varepsilon_c(\bm q_c)$ where $\bm q_c$ is the Bloch wave vector at the centre of the wave packet, while the gradient correction should be worked out \cite{GaneshSundaram1999}. In the presence of an external magnetic field, the expectation value of the Hamiltonian Eq. \eqref{Eq:semiclassicalapproach} with respect to the wave packet gives the energy 
$\varepsilon= \varepsilon_{c} - \bm m\cdot \bm B$, where $\bm m=\bm m(\bm k)$ is the \textit{intrinsic} orbital magnetic moment of the electron at each $\bm k$
\cite{GaneshSundaram1999,MingCheChang2008,XiaoDi2010}. The intrinsic OMM reads:
\begin{equation}
\label{Eq:IntrinsicOMM}
\bm m^{nn}(\bm k)=\frac{ie}{2\hbar}\langle \nabla_{\bm k} u^{n}_{\bm k}|\times (\varepsilon^{n}_{c} - H_0) | \nabla_{\bm k} u^{n}_{\bm k} \rangle.     
\end{equation}
The semiclassical theory also revealed the existence of a Berry phase correction to the density of states \cite{XiaoDi2005, XiaoDi2010}, 
which takes the form $\mathcal{D}(\bm k)=(2\pi)^{-d}(1+e \bm B \cdot \bm \Omega/\hbar)$ where $d$ is the spatial dimension of 
the system and $\bm \Omega^{nn}_{\bm k}=i\langle \nabla_{\bm k} u^{n}_{\bm k}|\times | \nabla_{\bm k} u^{n}_{\bm k} \rangle $ is the Berry curvature for band $n$. The equilibrium magnetisation density is calculated from the total energy as $\bm M_{ob}=-\partial \langle \varepsilon \rangle /\partial \bm B$ where the total energy $\langle \varepsilon \rangle = \int d\bm k \mathcal{D}(\bm k)[\varepsilon_{c}(\bm k)- \bm m (\bm k)\cdot \bm B]$, resulting in
\begin{equation}
\label{Eq:totalOMsemiclassic}
\bm M_{ob}= \int \frac{d\bm k}{(2\pi)^d} \left\{  \bm m(\bm k) + \frac{e}{\hbar}\bm \Omega_{\bm k} [\mu - \varepsilon_{c}(\bm k)] \right\}. 
\end{equation}
Here, $\mu$ is the chemical potential. This allows us to define a modified orbital magnetic moment at each $\bm k $ as
\begin{align}
\label{Eq:FullOMM}
\bm m^{nn}(\bm k) 
&\rightarrow
\frac{ie}{2\hbar}\langle \nabla_{\bm k} u^{n}_{\bm k}|\times (2\mu - \varepsilon^{n}_{c} - H_0) | \nabla_{\bm k} u^{n}_{\bm k} \rangle. 
\end{align}
The intrinsic orbital magnetic moment, Eq.~\eqref{Eq:IntrinsicOMM}, is associated with the self-rotation of the wave packet around its center of mass. 
The additional term leading to Eq.~\eqref{Eq:FullOMM} is the Berry phase correction associated with the motion of the center of mass of the wave packet in the semiclassical theory. In this context, the notion of a Berry phase correction to the density of states has been re-examined in Ref.~\cite{ValetRaimondi2023}, who pointed out that, rather than a band-dependent modification of the density of states in momentum space, this correction can be understood as spectral weight transfers induced by the Berry curvature, in other words a modification to the structure of the spectral weight function. 

\begin{figure}[tbp]
\centering
\includegraphics[height=3.5cm]{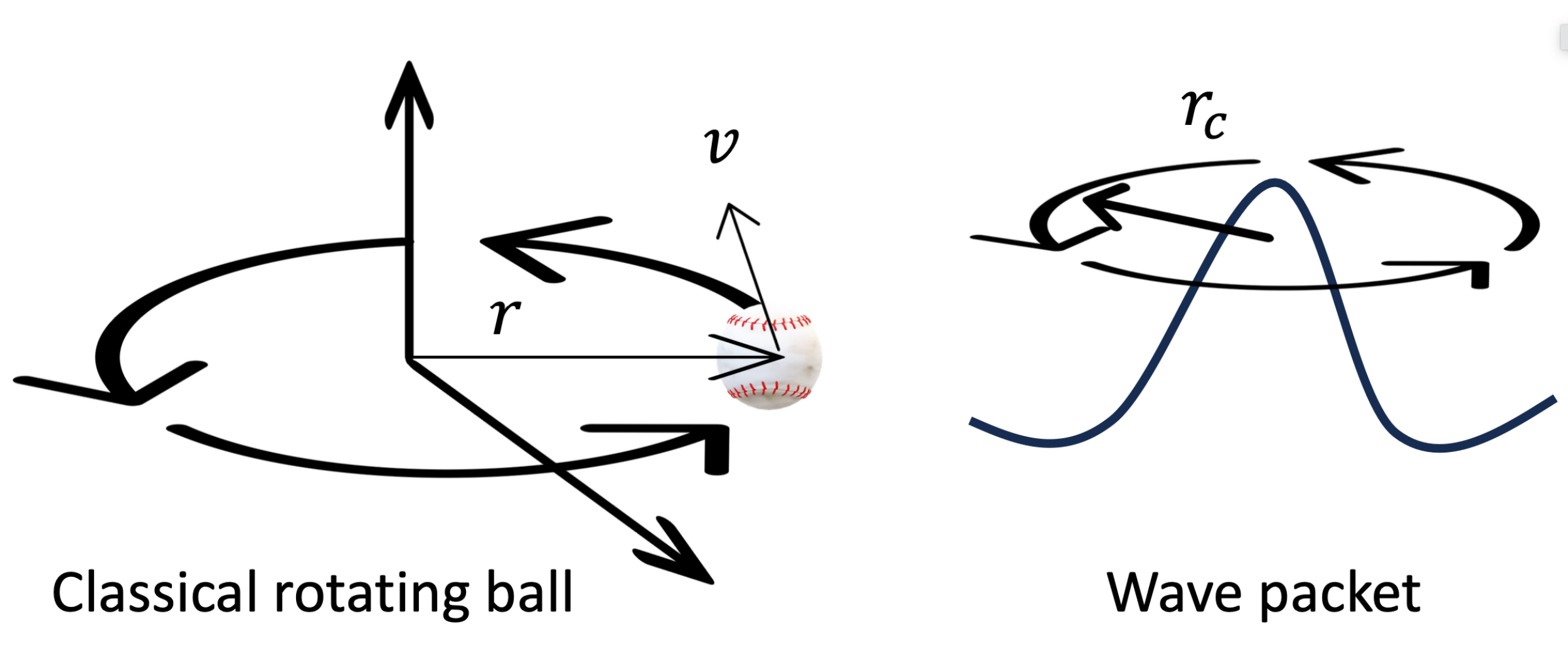}
\caption{Schematic depicting a rotating classical particle and `classical' (electron) wave packet rotating around its center. Note that the angular momentum $\bm L =\bm r\times \bm v$ depends on the point of reference. The semiclassical approach 
sets the center of the wave packet as the point of reference with respect to which $\bm L$ is measured. The orbital angular momentum of the electron wavepacket is related to its orbital magnetic moment as $\bm m =-e\bm L$, with $-e$ being the electronic charge.}
\label{Fig:wavepacketrotation}
\end{figure}

A formal derivation of the relationship between the OAM and the Berry curvature can be found in Ref.~\cite{SongJustin2019}. At each $\bm k$, the Berry curvature can be written 
as the cross product of the position operator $\bm r$ with itself, namely, $\bm \Omega^{nn}(\bm k) =i [\bm r \times \bm r]^{nn}_{\bm k}$, which is finite 
as long as we project onto a single band. 
Writing the velocity operator $\bm v=\frac{i}{\hbar}[H_0,\bm r]$ explicitly in the OAM we find,   
\begin{equation}
\label{Eq:relationOAMandBC}
\bm L^{nn}(\bm k)= \frac{i}{\hbar} \sum_{n'\neq n}(\varepsilon^{n}_{\bm k} - \varepsilon^{n'}_{\bm k}) [\bm r \times \bm r]^{nn}_{\bm k}.   
\end{equation}
Here \textit{only} the band off-diagonal elements of the velocity operator are included. This result suggests that, even though Bloch electrons are translationally invariant and delocalised, they can have a nonzero orbital angular momentum which, in equilibrium, comes from virtual couplings to other bands \cite{Culcer2005}. Xiao \textit{et al} used an alternative technique to derive the OAM that includes finite temperature~\cite{XiaoDi2006}. Another approach relies on thermodynamic potentials, which avoids the ill-defined position operator in periodic systems. This was considered in Ref.~\cite{ShiJunren2007}, which also extended the theory to include electron-electron interaction. Electron-electron interactions were also considered in Ref.~\cite{Aryasetiawan2016} using a Green's function approach.

\subsubsection{Wannier picture}

In order to overcome the problem of the unbounded position operator, it is possible to use Wannier functions $| \psi_{i} \rangle $. This was done in Ref.~\cite{Thonhauser2005} and extended in Ref.~\cite{CeresoliDavide2006} for multi-band systems and to include metals and Chern insulators, focussing on zero temperature. In this approach, the magnetisation is separated into two contributions, namely $\bm m^{I}_{i} =-\frac{e}{2} \langle \psi_{i}|(\bm r- \Bar{\bm r}_i)\times \bm v |\psi_{i} \rangle $
and $\bm m^{II}_{i}=-\frac{e}{2} \Bar{\bm r}_i\times \langle \psi_{i}| \bm v | \psi_{i} \rangle$,
with the centre of the Wannier function given by $\Bar{\bm r}_i$ and the total orbital magnetisation density is $\bm M = \sum_{i} (\bm m^{I}_{i}+\bm m^{II}_{i})$. Explicit calculation \cite{Vanderbilt2018} gives 
\begin{equation}
\label{Eq:contributionIwannierpicture}
\bm M_{I}=\frac{e}{2\hbar} {\rm Im} \int \frac{d\bm k}{(2\pi)^{d}} \langle \nabla_{\bm k} u^{n}_{\bm k}|\times H_0  | \nabla_{\bm k} u^{n}_{\bm k} \rangle
\end{equation}
and
\begin{equation}
\label{Eq:contributionIIwannierpicture}
\bm M_{II}=-\frac{e}{2\hbar}\int \frac{d\bm k}{(2\pi)^{d}}\varepsilon^{n}_{\bm k}\Omega^{nn}_{\bm k}.
\end{equation}
This approach provides the same results as the semi-classical method \cite{XiaoDi2005, XiaoDi2006}, specifically, Eq.~\eqref{Eq:contributionIwannierpicture}+Eq.~\eqref{Eq:contributionIIwannierpicture} 
= Eq.~\eqref{Eq:totalOMsemiclassic}.
The $\bm M_{II}$ contribution can be interpreted as the circulation of the center of mass of the wave packet. The $\bm M_{I}$ contribution is related to the self-rotation of the wave packet around its center of mass. Note that the self-rotation contribution is derived semi-classically from the gradient correction in Eq.\eqref{Eq:semiclassicalapproach}. In a magnetic field, the first term in Eq.\eqref{Eq:semiclassicalapproach} is a function of the gauge invariant crystal momentum, namely, $H_c=H_{c}(\bm q_c+\frac{e}{\hbar}\bm A)$. This can be expanded up to first order in the gauge potential and using the symmetric gauge $\bm A=\frac{1}{2}\bm B\times \bm r_c$ we get $H_c \approx H_{c}(\bm q)+\frac{e}{2\hbar} (\bm r_{c}\times \nabla_{\bm q} H_{c}(\bm q))\cdot \bm B$, that looks like the $\bm M_{II}$ contribution described in the Wannier function approach and offers the intuitive interpretation inherited from the semi-classical approach \cite{Fuchs2010}. In the context of the Wannier states, it is associated with a current (circulating) at the boundaries, i.e., the DOS contribution is linked to an edge effect \cite{Vanderbilt2018}.

In Ref.~\cite{Thonhauser2005}, the contribution $\bm m^{I}_{i}$ is referred to as local while $\bm m^{II}_{i}$ is denoted as itinerant. The term \textit{local} is to be understood as atomic, in the sense that Bloch bands are ultimately related to atomic wave functions, and atomic orbitals beyond $s$ carry a finite angular momentum. Depending on the orbitals involved in a particular band, electrons in that band may inherit a finite angular momentum from the parent atomic orbitals. The term itinerant, on the other hand, is associated with hopping between different atoms. 

Although the end result is the same in the Bloch and Wannier pictures, individual terms cannot always be placed in exact correspondence. One formulation that seeks to encapsulate these correspondences for an equilibrium system is Eq.~(18) of Ref.~\cite{XuanQuekPRR2020}. Further insight into the transition between the Wannier and Bloch pictures can be gained by considering the extreme case in which the solid is a periodic array of non-overlapping atoms which carry a well defined orbital moment. In this case the macroscopic magnetization equals the atomic orbital moment over the cell volume, and the standard formulas of the \textit{modern theory} given above contain all of the atomic-OAM \cite{RestaRaffaele2020}.
 
\subsection{Centrosymmetric systems}

The seminal paper introducing the orbital Hall effect \cite{Bernevig2005} relied on a centro-symmetric approximation to the valence band of Si, a $T_d$-semiconductor, which was recently revisited in the fully numerical study of Ref.~\cite{BaekInsu2021}. Using ${\bm k}\cdot{\bm p}$ theory, the carrier wave function in the valence band is expanded in terms of Bloch states at the valence band edge, which are associated with atomic orbitals of $p$-symmetry, carrying a finite OAM $L = 1$. Hence, the electrons possess an OAM inherited from the lattice.

Spin-orbit coupling is neglected in  Ref.~\cite{Bernevig2005}, such that the OAM is a good quantum number at the band edge ${\bm k} = 0$. In the ${\bm k}\cdot{\bm p}$ approximation, the atomic OAM is contained in the lattice-periodic wave functions at the band edge. The wave functions at finite wave vector mix states with different OAMs, so that the OAM of itinerant electrons away from the band edge is not conserved. The effective ${\bm k}\cdot{\bm p}$ Hamiltonian can be written in terms of effective orbital angular momentum operators, which correspond to the OAM of the parent atomic orbitals. The Berry curvature in these effective models is zero, and the OAM found using the modern theory is also zero. The Hamiltonian thus obtained is similar to the more familiar spin-orbit Hamiltonians. However, the interaction couples the OAM and the wave vector rather than the spin and the wave vector. The finite carrier momentum away from the band edge tilts that OAM and the good quantum number for bulk Si holes is ${\bm L}\cdot{\bm k}$. Out of equilibrium, the finite net momentum results in OAM transport in the same way as in the spin-Hall effect. Since the quantity calculated is in principle not conserved, Ref.~\cite{Bernevig2005} identifies a conserved OAM operator, although the applicability of this conserved OAM beyond ${\bm k}\cdot{\bm p}$ models requires further investigation.

It is possible that the simplified description of Ref.~\cite{Bernevig2005} does not correspond to realistic Si hole samples. Tight binding studies indicate that an effective $8 \times 8$ model including $s$ and $p$ orbitals is the minimum requirement for reproducing the band structure to a reasonable degree of accuracy \cite{Papaconstantopoulos1980}. In addition, spin-orbit coupling in Si is weak but not negligible, and the full Luttinger model needs to be used for an accurate description of the physics. Indeed, recent years have revealed that the spin-orbit coupling of Si holes plays a leading role in electrically operated quantum computing platforms and gives rise to phenomena such as coherence sweet spots \cite{Silvano2022, Fang2023, Zhanning2024}. When spin-orbit coupling is included the Berry curvature at each ${\bm k}$ does not vanish even in the spherical Luttinger model, and neither does the OAM calculated using the modern theory. 

Nevertheless, starting from the idea of Ref.~\cite{Bernevig2005}, it was shown that the ${\bm k}\cdot{\bm p}$ Hamiltonians of many inversion symmetric systems can generically be written in terms of effective OAM operators and powers of the wave vector. Using powerful symmetry arguments similar to those presented for Si, the atomic-OAM has been studied in various inversion symmetric systems \cite{JoDaegeun2018, SeungynHan2022, ChoiYoungGwan2023}, in which the Hamiltonian can be written in terms of an effective orbital angular momentum operator $L$. The connection between this operator and the microscopic angular momentum of the modern theory requires further investigation. Nevertheless, the important role played by hybridisation of atomic orbitals has been highlighted. Hybridisation results in a non-trivial orbital texture and a corresponding coupling between the orbital degrees of freedom and the momentum \cite{GoDongwook2018, JoDaegeun2018, DongwookGo2021, BuschOliver2023}, which leads to an orbital magnetisation even when the angular momentum is quenched in equilibrium. Such hybridisation was referred to as an \textit{orbit-orbit} interaction in Ref. \cite{Bernevig2005}.

\section{Theory of Non-equilibrium OAM}
\label{Sec:nonequilibriumeffects}

A non-equilibrium OAM density can be generated via the orbital magneto-electric effect (OME) \cite{YodaTaiki2015, YodaTaiki2018, CysneTarik2023} and the orbital Hall effect (OHE), itinerant \cite{Tokatly2010, BhowalSayantika2020, BhowalSayantika2020-III, PezoArmando2023} and atomic, \cite{Bernevig2005, JoDaegeun2018, GoDongwook2018, PezoArmando2022, CysneTarik2022} which are the two mechanisms for charge-to-orbital current conversion. The OME represents the generation of a steady-state non-equilibrium OAM density by an electric field throughout the bulk of a sample. The OHE is an OAM current that flows transversely to an applied electric field and results in an OAM accumulation at the edge of the sample. Both effects can occur in time-reversal ($\mathcal{T}$) invariant and time-reversal breaking systems, although the focus has been on the former till now. The OME is restricted to gyrotropic systems: inversion symmetry ($\mathcal{P}$) breaking is a necessary but not sufficient condition. Gyrotropic crystals are characterised by an asymmetric permittivity tensor, e.g. $\varepsilon_{xy} \ne \varepsilon_{yx}$, and left and right elliptical polarisations can propagate at different speeds. Only certain non-centrosymmetric crystal classes exhibit gyrotropic point group symmetry \cite{CrystallographyBook,AuthierBook2003}. The OHE is not restricted by any crystal or time reversal symmetry considerations. 

Non-equilibrium effects can be classified as intrinsic and extrinsic depending on whether they originate in the band structure or disorder. Intrinsic responses to an electric field are associated with geometrical effects encoded in the band structure of the system. In contrast, extrinsic effects are connected to disorder, although they can be of zeroth order in the disorder strength. This distinction is important because, depending on the scaling of the disorder, extrinsic contributions can be allowed depending on the system's combined $\mathcal{P}\mathcal{T}$ symmetry. Consider the linear response of a given operator $\mathcal{O}$ to an electric field. The average of such an operator can be written as $\langle \mathcal{O} \rangle=\Lambda E$, where $\Lambda$ is the corresponding response function and $E$ is the electric field. Typically, the response coefficient can be written as $\Lambda = \Lambda^{(-1)} + \Lambda^{(0)}$ where the first term is inversely proportional to the disorder strength, and the second one is independent of disorder strength. Equivalently, $\Lambda^{(-1)}$ is proportional to a relaxation time $\tau$ that acts as a measure of the disorder strength, which in the first Born approximation is inversely proportional to e.g. the impurity concentration, $\tau\sim n^{-1}_{imp}$.

In general, the response $\Lambda^{(0)}$ can be sub-classified into two parts: one that is related to the geometry of the band structure and that we will call intrinsic and one that scales as a ratio of two different relaxation process which are parametrically different with respect to the Fermi energy but are both linear in impurity density, making the overall response formally independent of impurity concentration. We will refer to it as the extrinsic response of zeroth order in disorder or disorder correction to the intrinsic response. This classification is well known in charge and spin currents \cite{Inoue2003, Inoue2004, Inoue2006, Culcer2017, BurgosPRR2022}. It is expected to be the case in linear response, although disorder corrections to OAM dynamics have received comparably little attention. So far, the intrinsic channel has been studied the most because it can be implemented in {\it ab-initio}  calculations \cite{LopezMG2012, MarzariNicola2012, CeresoliDavide2010, SalemiLeandro2022, CostaMarcio2023} in order to test properties of the band structure. In this theoretical section we will consider the OME and OHE, their intrinsic and extrinsic contributions in connection to $\mathcal{P}$ and $\mathcal{T}$ symmetries, and their applications in orbitronics.

\subsection{Orbital magneto-electric effect}
\label{Sec:OEE}

The OME is a net OAM density generated in the bulk of a sample by an applied electric current. In its simplest form, it can be understood as follows. The charge carriers in the system possess an OAM. In the presence of $\mathcal{T}$-symmetry, the net OAM density in equilibrium vanishes. When an electric field is applied, the Fermi surface is shifted away from equilibrium, and the net OAM density can be nonzero -- this is allowed if the system has gyrotropic symmetry. The induced OAM density is proportional to the electric field and the Fermi surface shift. The latter depends on the effective scattering time $\tau$, which represents dissipation and guarantees $\mathcal{T}$-breaking. This mechanism, depending on the Fermi surface and disorder and involving band-diagonal matrix elements of the OAM, is also known as the orbital Edelstein effect \cite{YodaTaiki2015, YodaTaiki2018, ChengwangNiu2019, HeWenYu2020-II}, and is the counterpart of the spin Edelstein effect  \cite{EDELSTEIN1990} or inverse spin galvanic effect \cite{Ganichev2002} from semiconductor spin dynamics.

\begin{figure}[tbp]
\centering
\includegraphics[width=\linewidth]{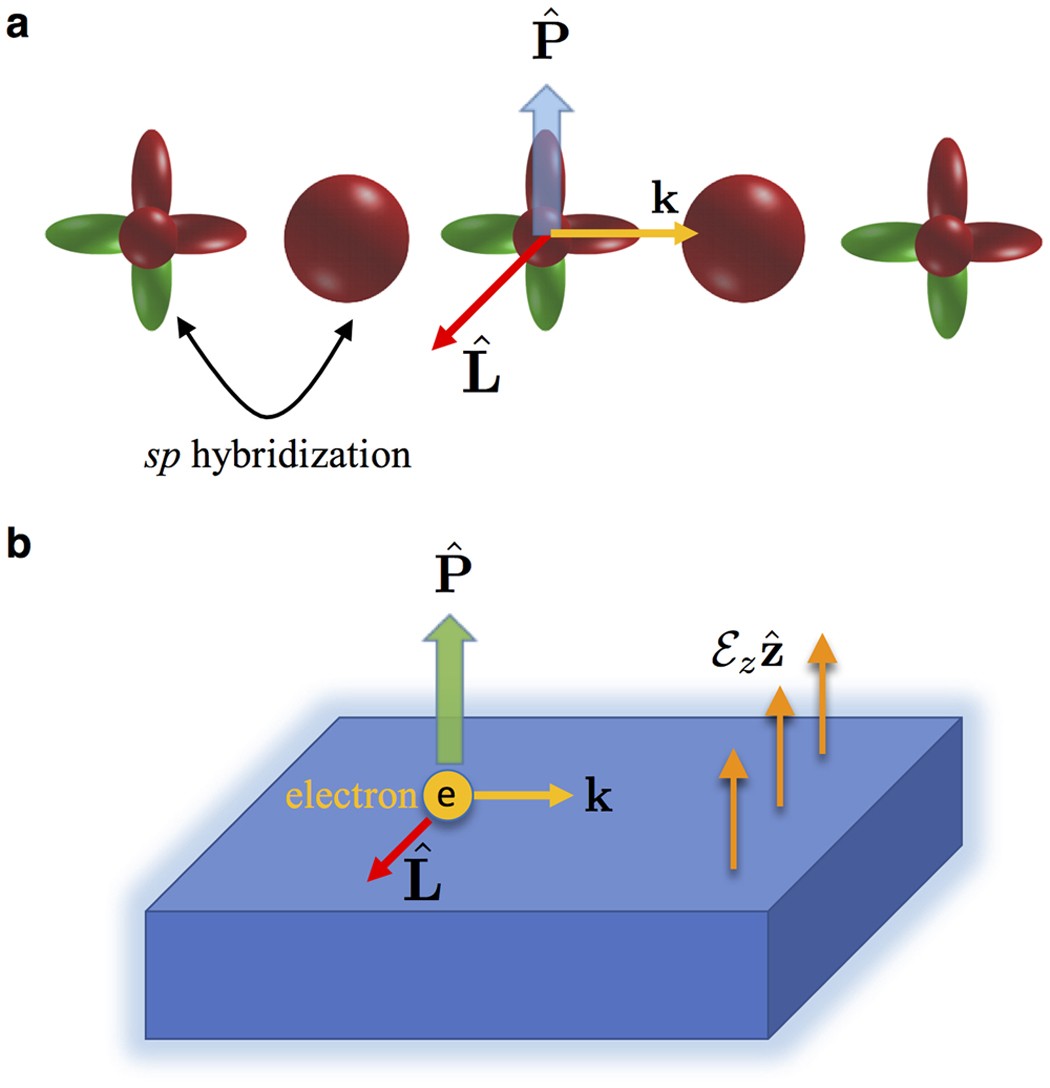}
\caption{ Figure taken from Ref.~\cite{GoDongwook2017}. a) the system can be prepared 
in a non-vanishing orbital angular momentum state and the hybridization generates an
electric polarization $\hat{\bm P}$. b) The coupling between this electric polarization 
and the surface gradient can lead to the Rashba-like orbital coupling, which leads to the OME. }
\label{Fig:atomicorbitalME}
\end{figure}

Interestingly, this orbital effect was predicted over a decade ago in zeolite-templated carbon systems ~\cite{Koretsune2012}. To generate an OME inversion symmetry can first be broken by orbital hybridization \cite{GoDongwook2017} or by a gate electric field \cite{GoDongwook2021}. 
We exemplify this in Fig.~\eqref{Fig:atomicorbitalME}.
The OME has also been denoted as the kinetic magneto-electric effect in $\mathcal{T}$-symmetric systems, and was found by the authors of Ref.~\cite{OsumiKen2021} to be exceedingly strong in topological insulators. The OME can also be intrinsic in $\mathcal{T}$-breaking systems (which preserve the combined $\mathcal{PT}$-symmetry) \cite{ShinadaKoki2023}, in which case it involves band off-diagonal matrix elements of the OAM and virtual inter-band transitions induced by the electric field, and are a form of inter-band coherence~\cite{Culcer2017}. This intrinsic form of OME does not involve dissipation and is a Fermi sea response. A similar effect to the OME, also occurring in gyrotropic systems, is the gyrotropic magnetic effect (GME)\cite{ZhongShudan2016}, which represents a current density driven by a slowly varying magnetic field and involves both the spin and the orbital magnetic moments. Finally, although the OME does not occur in inversion symmetric systems, a non-linear OME can exist \cite{baek2024nonlinear}. 

The OME was predicted by Yoda et al. in Ref.~\cite{YodaTaiki2015} in chiral crystals with time-reversal symmetry ($\mathcal{P}\mathcal{T}$-broken) based on their sizable OAM. The authors studied a generic model, predicting its applicability to materials such as Se and Te, which are helical with broken inversion symmetry. The hopping integral determines the size of the effect. In a similar system with helical structure (in the hopping integral to break inversion symmetry) and spin-orbit coupling, both spin and orbital magneto-electric effects are predicted to coexist \cite{YodaTaiki2018}. The model was extended to include more general materials with $\mathcal{P}\mathcal{T}$-broken symmetry in Ref.~\cite{HaraDaisuke2020}. Here, a toy model was introduced to mimic a 3D polar metal. The current-induced orbital magnetisation results from an electronic current forming closed loops, similar to a classical solenoid. For SnP in its polar phase, an OME of $\sim 0.63$G is predicted for an electric field $E_x=10^4$V/m and a life time of $\tau=1$ps.

A recipe for producing a strong OME is to start with a material with a strong OAM and lower the symmetry to enable the OME. One example is provided by TMDCs, such as NbS or NbSe, which are $\mathcal{T}$- invariant and break $\mathcal{P}$, but C$_{3z}$ symmetry makes them non-gyrotropic. Bhowal et al. \cite{BhowalSayantika2020-II} showed that gyrotropy can be engineered by lowering the symmetry to C$_{2v}$ via uni-axial strain. The strained TMDC is a $\mathcal{P}\mathcal{T}$-broken state and exhibits the OME. The authors termed the effect orbital gyrotropic magneto-electric effect to emphasize the importance of the gyrotropic symmetry induced by strain. Its sign can be tuned using tensile or compressive strain. Similarly, Ref.~\cite{HeWenYu2020-II} showed that strain drastically enhances the OME in twisted bilayer graphene.

The reduction from C$_{3z}$ to C$_{2v}$ symmetry via geometry rather than strain in order to allow an electric-current-driven OAM density was also theoretically considered in Ref.~\cite{CysneTarik2021-I} for $p$-band systems and in Ref.~\cite{CysneTarik2023} for TMDCs. Ref.~\cite{CysneTarik2023} considered monolayers (L) and bilayers (2L) of MoS$_2$ nanoribbons with zigzag edges and analysed the magnetisation profile at the edges. By symmetry, the OME is not allowed in 2L-MoS$_2$ because $\mathcal{P}$-symmetry is restored, but is allowed in L-MoS$_2$. 

Considering the orbital moment operator $L_i$, in linear response, its average can be written as $\langle L_i \rangle =\chi_{ij} E_j$, where $\chi_{ij}$ is the response coefficient. The linear response coefficients can be classified according to their dependence on the disorder strength, as $\chi_{ij}=\chi^{(0)}_{ij}+\chi^{(-1)}_{ij}$, where the superscript refers to the order in disorder strength, as discussed above. 
We recall that the OME requires broken $\mathcal{P}$ symmetry. As a result, if the system breaks $\mathcal{T}$, then it is $\mathcal{P}\mathcal{T}$-symmetric by definition and the contribution $\chi^{(0)}_{ij}$ is the only allowed response. For instance, the intrinsic OME in metals was considered in Ref.~\cite{XiaoCong2021-I, XiaoCong2021-II} driven by an electric field and by a temperature gradient and in Ref.~\cite{ShinadaKoki2023} at finite temperature. In particular, Ref.~\cite{XiaoCong2021-II} used the semi-classical approach and derived equations that solely depend on band structure as a signature of intrinsic effects. They applied their theory to CrI$_3$ van der Waals layered magnetic material, where both $\mathcal{P}$ and $\mathcal{T}$ are broken. They predicted larger values of OME as compared to the Rashba Edelstein effect \cite{Johansson2016}.

Non-centrosymmetric antiferromagnets CuMnAs and Mn$_2$Au represent $\mathcal{P}\mathcal{T}$-symmetric states where a non-equilibrium OAM density is allowed because of broken inversion symmetry. This was studied by first principles in Ref.~\cite{SalemiLeandro2019}. The authors also included spin-orbit coupling so that both the spin Edelstein effect and OME are present. The orbital channel dominates the overall magnetisation, although the dominance is more pronounced in CuMnAs than in Mn$_2$Au. The authors showed a Rashba-like symmetry in the OAM density of atomic orbitals created by the external electric field (meaning atomic contribution) and, more importantly, that the induced orbital texture 
strongly depends on the N\'eel vector orientation. This can provide a way for N\'eel vector manipulation in linear response in addition to the nonlinear regime based on the Hall effect \cite{ShaoDingFu2020, WangChong2021}.

The OME can also occur on insulator surfaces and interfaces. In particular, it is predicted for Chern and Z$_2$ topological insulators \cite{OsumiKen2021} to be larger than in metals. This was theoretically predicted in the 2DEG at SrTiO3 (STO) interfaces, which do not exhibit magnetic order in their ground state \cite{JohanssonAnnika2021}. The spin/OME co-exist in this system, but the OME is expected to exceed its spin counterpart by more than one order of magnitude. X-ray magnetic circular dichroism (XMCD) can separate the spin and orbital contributions. 

$\mathcal{P}\mathcal{T}$-broken systems can have a nonzero Berry curvature dipole \cite{Sodemann2015, MaQiong2019}. Since both the OME and the Berry curvature dipole are present in $\mathcal{P}\mathcal{T}$-broken systems and depend on intrinsic geometric quantities, a connection between them was proposed in the case of strained transition-metal dichalcogenides (TMDC) \cite{SonJoolee2019}, where the OME was expressed as a Berry curvature dipole. This is also the case for LaAlO$_3$/SrTiO$_3$~\cite{Lesne2023}.
With the Berry curvature dipole density denoted by ${\bm D}$, 
the magnetisation density $\bm M \sim (\bm D \cdot \bm E)\hat{\bm z}$, and a non-linear current ${\bm J}_{Hall}\sim (\bm D \cdot \bm E)\hat{\bm z}\times \bm E$ is also present, as shown in Ref. \cite{SonJoolee2019}. The non-linear current can be interpreted as a consequence of the Lorentz force due to an effective magnetic field $\propto (\bm D \cdot \bm E)$.

The OME can also occur in superconductors. In this context, we first recall that a spin-Edelstein effect was predicted in polar superconductors with Rashba spin-orbit coupling~\cite{Edelstein1995} and in $d$-wave superconductors~\cite{IkedaYuhei2020}, while magnetisations in different gyrotropic superconductors were addressed in~\cite{HeWen-Yu2020}. A spin supercurrent can be generated by injecting a charge current into a Josephson junction with heavy metals and FM~\cite{BergeretTokatly2015,LinderJacop2017}. Some experiments support the applicability of superconducting plus HM devices \cite{Wakamura2015}. The generation and manipulation of spin currents in superconducting devices is very promising for applications since it may be dissipationless \cite{HeJames2019}.

It is natural to expect both spin and orbital magnetisation in response to a supercurrent in superconductors with gyrotropic symmetry. Such an orbital magnetisation was predicted in chiral $p$-wave superconductors \cite{Annett2009, Robbins2020}. For superconductors with an orbital Rashba-like interaction, an intrinsic OME was predicted in ~\cite{ChirolliLuca2022} expected to be at least one order of magnitude greater than the spin Edelstein effect. Another interesting manifestation of the OME occurs in Moir\'e materials \cite{Twistnoics-Allan, JianpengLiu2021}. 
For instance, in twisted bi-layer graphene (TBG) with a twist of $\theta = 1.1^\circ$ a superconducting state is created \cite{Cao2018}. The system becomes strongly correlated at this special \textit{magic angle}. This is a consequence of the creation of Moir\'e patterns and the appearance of flat bands that reduce the kinetic energy such that the electron-electron correlation becomes the dominant interaction. Recent experiments reported the anomalous Hall effect in bilayer graphene \cite{XiaoboLu2019, SharpeAaron2019, Serlin2020, Chen2020, Sharpe2021, TschirhartCL2021} indicating ferromagnetic behaviour having an orbital origin because the spin-orbit coupling is negligible. Berry curvature effects dominate the magnetism, which produces a non-zero Chern number. Materials with these characteristics are called \textit{orbital Chern insulators}. Intuitively, TBG's orbital magnetization can be considered a current loop in the Moir\'e super-lattice.
This can give rise to anti-ferromagnetic states for two counter-rotating current-loop at different valleys \cite{LiuJianpeng2021} or to ferromagnetic states due to current loop at the boundaries of the Moir\'e super-cells \cite{LiuJianpeng2019}. Magic angle twisted bilayer graphene (MATBLG) is advantageous because it offers an easy way of electrical control of orbital magnetisation as reported in Ref.~ \cite{ZhuJihang2020, Polshyn2020, HeWenYu2020-II}. In Ref.~\cite{HeWenYu2020-II}, the authors demonstrated that the OME generated by an electric field in MATBLG could flip the spontaneous orbital magnetisation induced in the same material by strong electron correlations.

\begin{figure}[tbp]
\centering
\includegraphics[width=\linewidth]{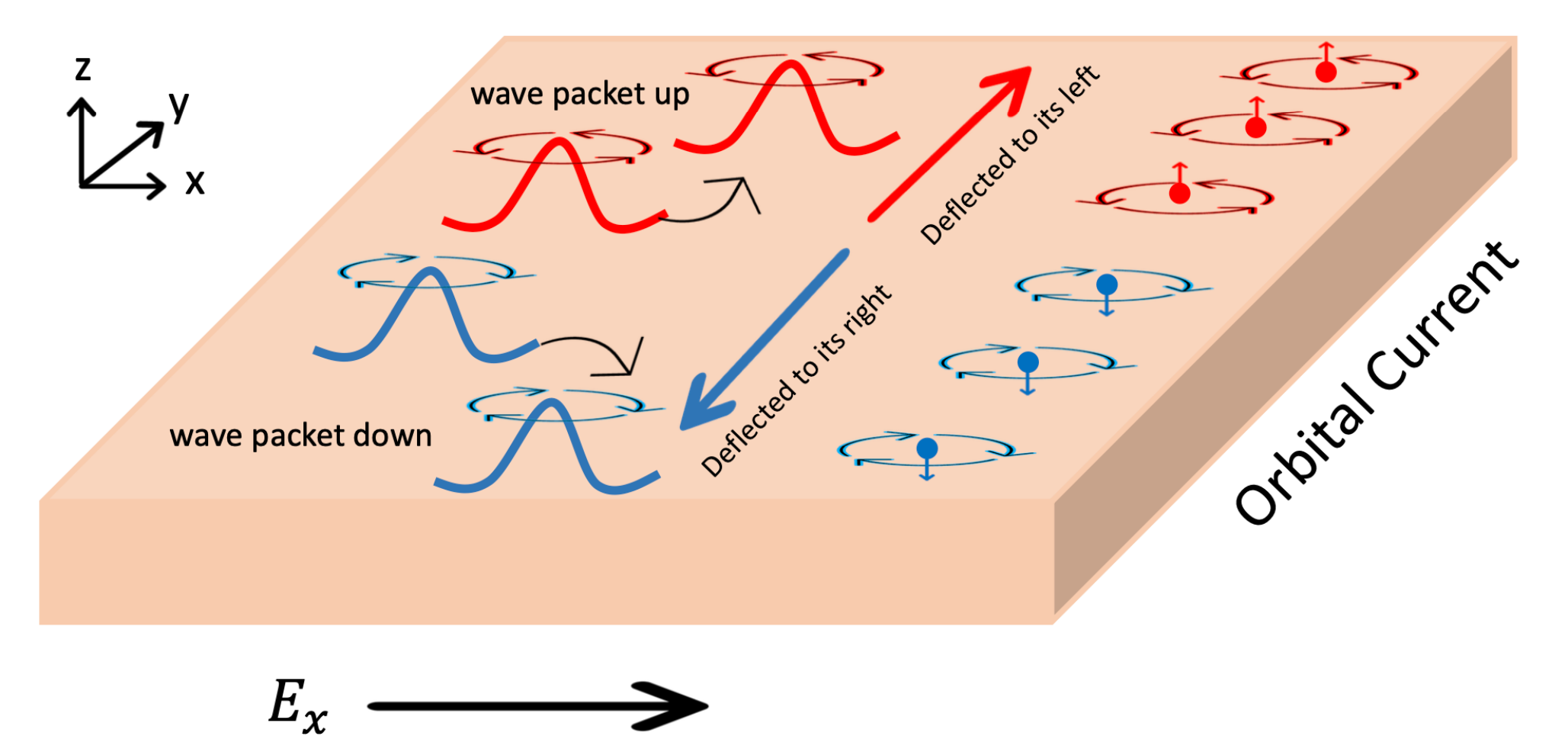}
\caption{Sketch of the orbital Hall effect due to intrinsic orbital angular momentum. Wave packets with different angular self-rotations can be deflected in different directions when an electric field is applied. This gives rise to an orbital current without charge current.}
\label{Fig:orbitalHalleffect}
\end{figure}

\subsection{Orbital Hall effect}
\label{Sec:OHE}

\begin{figure}[tbp]
\centering
\includegraphics[width=\linewidth]{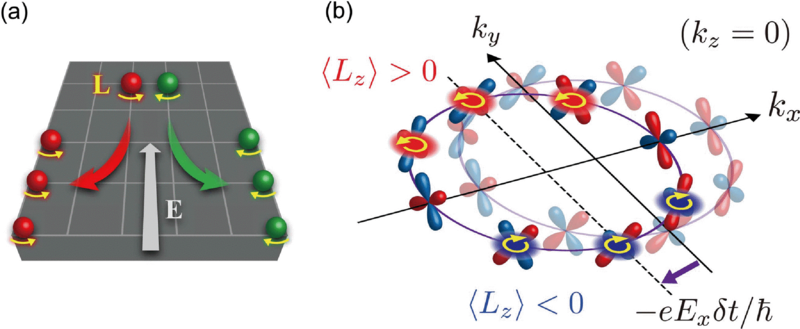}
\caption{Figure taken from Ref.~\cite{JoDaegeun2018}. a) sketch of electron with opposite OAM moving in opposite direction in order to generate OHE. b) The average $\langle L_z \rangle$ is opposite for
opposite $k_y$, then the OHE arises. 
}
\label{Fig:atomicorbitalHalleffect}
\end{figure}

The orbital Hall effect is the flow of OAM \textit{transversal} to an applied electric field, as sketched in Fig.\eqref{Fig:orbitalHalleffect}. The
transversal current leads to an accumulation of OAM at the boundaries of the material. It was originally predicted in $p$-doped Si \cite{Bernevig2005}, followed by $d$-transition metals \cite{Kontani2009, Tanaka2008, Kontani2008} and in $f$-electron systems \cite{Tanaka2010}. More recently, the OHE has been proposed as the origin of the valley-Hall effect \cite{BhowalSayantika2021} and predicted to be strong in graphene and TMDCs \cite{BhowalSayantika2020, BhowalSayantika2020-III}. We illustrate one possible atomic mechanism for the OHE in Fig.~\eqref{Fig:atomicorbitalHalleffect}.

The orbital current operator is defined as $\mathcal{J}^{i}_{j}=\{v_{i}, L_{j} \}/2$. For out-of-plane OAM, we can consider in linear response the averaged orbital current written as $\langle \mathcal{J}^{i}_{z} \rangle = \alpha_{ij}E_{j}$. As before, $\alpha_{ij}=\alpha^{(0)}_{ij} + \alpha^{(-1)}_{ij}$, depending on the scaling behaviour with disorder, where $\alpha^{(0)}_{ij}$ contains the intrinsic dissipationless OHE together with some disorder corrections discussed below, while $\alpha^{(-1)}_{ij}$ is responsible for dissipation. Most of the work on the OHE has focussed on the intrinsic channel of the susceptibility $\alpha^{(0)}_{ij}$. For simplicity, let us consider that only the band diagonal elements of the orbital moment
$L_{z}$ are relevant for transport. Then, the intrinsic channel of the orbital current follows as
\begin{align}
\label{Eq:generalexpressionOHE}
\langle \mathcal{J}^{\alpha}_{z} \rangle 
&=\frac{ eE_b }{\hbar} \sum_{{\bm k}; n\neq n'} ( L^{n'n'}_z + L^{nn}_z ) {\rm Im}[\mathcal{Q}^{nn'}_{b\alpha}]
n_{FD}(\epsilon^{n}_{\bm k}),
\end{align}
where $\mathcal{Q}^{nn'}_{b\alpha}=\mathcal{R}^{nn'}_{\bm k,b}  \mathcal{R}^{n'n}_{\bm k,\alpha}$ is the quantum metric tensor, where the Berry connection is defined with the periodic part of the Bloch wave function as
$\mathcal{R}^{nn'}_{\bm k,j}=i\langle  u^{n}_{\bm k}| \partial_{k_j} u^{n'}_{\bm k}\rangle$.
We stress that in the above expression, we can use either the atomic orbital momentum operator for $L_z$ for atomic effect or
the matrix elements of the operator $L_i=\epsilon_{ijl}\{ r_j, v_l \}/2$ for the total 
contribution \cite{BhowalSayantika2020, PezoArmando2023}. Also, it should be noted that the geometrical interpretation of Eq. \eqref{Eq:generalexpressionOHE} 
is present in the imaginary part of the product of the two Berry connections, which leads basically to the Berry curvature. Moreover, Eq. \eqref{Eq:generalexpressionOHE} expresses 
a transversal flow of $( L^{n'n'}_z + L^{nn}_z )$ because the other terms are the well-known anomalous velocity leading to the anomalous Hall effect \cite{XiaoDi2010}. The connection between the anomalous velocity and the orbital magnetic moment in the presence of a magnetic field was studied in Ref.~\cite{KamalDas2021}. They showed that intrinsic thermo-electric and thermal Hall effects contributions arise from the orbital magnetisation contained in the self-rotation of a wave packet. 

The OHE was predicted around the same time as the spin-Hall effect (SHE). Much of the initial effort on the topic was devoted to distinguishing the OHE from the spin and anomalous Hall effects in Pt \cite{Kontani2007}, transition metals \cite{Tanaka2008, Kontani2009}, Sr$_{2}$Ru O$_{4}$ and related materials \cite{Kontani2008}, and heavy-fermion systems \cite{Tanaka2010}. Disentangling the OHE from the SHE was a significant task since spin ($\bm S$) and orbital angular momentum ($\bm L$) exhibit the same physics and generally co-exist, and in many cases we need to consider the total angular momentum $\bm J_{tot}= \bm L + \bm S $ \cite{Tanaka2010}. 
The coexistence of both effects relies on spin-orbit coupling (not necessarily strong), which converts OHE to SHE, enhancing the spin Hall conductivity
\footnote{The difference in the non-equilibrium susceptibility for both is that for spin, it is proportional to the correlation $C \sim \langle \bm L \cdot \bm S\rangle $ while for the orbital, it is proportional to the orbital quantum number \cite{Kontani2008, Kontani2009, Tanaka2008, Tanaka2010}.}.

The OHE has come to the fore again in recent years when large signals were predicted in centro-symmetric systems \cite{GoDongwook2018}, including light materials with weak spin-orbit coupling \cite{JoDaegeun2018}. These papers studied the coexistence of SHE and OHE \cite{DongwookGo2021} and showed that the OHE could be comparable or larger to the SHE. Ref.~\cite{BaekInsu2021} revisited the OHE in Si and Ge and $\alpha$-Sn and found that the OHE conductivity was large and negative. A recent \textit{ab initio} study identified ${\mathcal T}$-odd \textit{magnetic} OHE and SHE in ferromagnets \cite{SalemiLeandro2022-I}, which appear to be comparable or smaller than the usual OHE and SHE, which are ${\mathcal T}$-even. Another paper by the same authors \cite{SalemiLeandro2022} focussed on monatomic metals and found the OHE to be small in $sp$-metals but significantly larger in $d$-metals near the middle of the $d$-series. An orbital Hall insulator phase has been predicted in analogy with the spin-Hall insulator \cite{CanonicoLuis2020}. 

Orbital current injection into a ferromagnet leads to a torque on the magnetisation, termed the orbital torque \cite{Dongwook2020-I}, which can be enhanced around hotspots in momentum space \cite{GoDongwook2023}. The reciprocal effect, in which magnetisation dynamics in a ferromagnet acts as a source of OAM, is known as \textit{orbital pumping}. This has been considered recently from a theoretical perspective with Refs.~\cite{Dongwook2023orbitalPump, Seungyunhan2023} addressing this effect in a bilayer of NM/FM. Their predictions show important differences 
in the FM material. For 3$d$ FM, they showed that the effect is more prominent in bi-layer Ti/Ni than Ti/Fe, which appears to be in qualitative agreement with experimental studies \cite{hayashi2023}. Likewise, Ref.~\cite{Seungyunhan2023} studied the effect of lattice dynamics on OAM pumping.

The OAM has begun to receive considerable attention in 2D Dirac materials such as graphene and TMDCs \cite{BhowalSayantika2021, CysneTarik2021, PezoArmando2022, CysneTarik2022, PezoArmando2023, BuschOliver2023, CostaMarcio2023}. The principal motivation in this area was the argument by Bhowal and Vignale that the OHE is at the root of the observed \textit{valley Hall effect} (VHE) \cite{BhowalSayantika2021} in gapped graphene. While the VHE was conventionally understood as an anomalous Hall effect with a different sign in each valley, Bhowal and Vignale argued that the VHE fundamentally represents a flow of OAM. For the re-interpretation of this VHE as a \textit{valley orbital Hall effect}, it is helpful to examine Eq.\eqref{Eq:generalexpressionOHE}. The first observation is that the orbital current is a Fermi sea response. Second, the factor in parenthesis is the sum of the orbital moment in the conduction band and valence band, which are the same, and the rest of the elements of the equation constitute the anomalous velocity, which is transversal to the electric field. Then, such an equation represents a transversal flow of $L_{z}$. Finally, for two different valleys related by time-reversal symmetry, both the orbital moment and the transverse velocity change sign, meaning that opposite orbital moments (defined in opposite valleys) flow in opposite directions. This effect decreases with the size of the gap, which is a clear signature of an inter-band coherence effect, meaning that virtual transitions between conduction and valence bands dominate the mechanism. 

Depending on the material under consideration, both VHE and OHE can be present. Since both effects lead to a transversal flow of OAM, a distinction between these effects is important. As an instance, in the mono-layer of TMDC, both effects exist in the atomic contribution generating the non-trivial accumulation of orbital moment at the edges \cite{XueFei2020, CanonicoLuis2020, BhowalSayantika2020}. Cysne \textit{et al.} proposed a bi-layer of 2H-MoS$_2$  \cite{CysneTarik2021} as a candidate to single out the accumulation of OAM due to OHE since inversion symmetry is restored, and then the flow of OAM due to valley Hall effect vanishes. In such a system, the VHE can be induced by an external gate voltage whose role is to break inversion symmetry \cite{KamalDas2023}. 

In a generic sample, both the atomic and the itinerant contributions to the OAM are present, and these are beginning to be investigated on the same footing. Ref.~\cite{CysneTarik2022} made use of a low energy effective Dirac Hamiltonian for a transition metal dichalcogenide bilayer and compared two approaches to handling the OAM. One approach involves the intra-atomic approximation, which only accounts for intra-site contributions to the OAM, while the second accounts for both intra-site and inter-site contributions. The authors showed that the two approaches agreed qualitatively, both predicting a finite OHE within the gap, while exhibiting quantitative differences as expected. Interestingly, Pezo \textit{et al.} \cite{PezoArmando2022} predicted a somewhat different result, showing that the atomic contribution dominates in wide band-gap semiconductors such as MoS$_2$. However, the authors found the itinerant contribution dominates in narrow band-gap semiconductors such as SnTe or PbTe and transition metals. The gap parameter plays a vital role in dictating which contribution dominates. In a subsequent paper, the same authors showed that if the orbital texture induced by hybridization is quenched, then the itinerant contribution dominates \cite{PezoArmando2023}.

\subsection{Disorder in the OHE}

The past two decades have revealed the critical role played in transport by off-diagonal terms in the density matrix, which are associated with band mixing or inter-band coherence \cite{Culcer2017, BurgosPRR2022}. These terms often provide the main contributions to transverse transport phenomena such as the anomalous Hall, spin-Hall, and orbital Hall effects. Inter-band coherence effects can be intrinsic, driven by band structure mechanisms that may be topological, or extrinsic, driven by disorder. 

Importantly, disorder produces corrections to transport coefficients that are formally of zeroth order in the disorder strength and compete with the intrinsic contributions. The fundamental reason behind these corrections is the following. Transport theory requires the condition $\varepsilon_F \tau/\hbar \gg 1$ to be satisfied, where $\varepsilon_F$ is the Fermi energy and $\tau$ represents a characteristic momentum relaxation time that can be used as a measure of the disorder strength. In light of this, transport theory is formulated as an expansion in the small parameter $\hbar/(\varepsilon_F \tau)$. Nevertheless, the first term in this expansion is the term leading to the longitudinal Drude conductivity, which is $\propto \tau$. Hence, it is formally of order $(-1)$ in the small parameter. The appearance of this term is understood on physical grounds -- it represents the shift in the Fermi surface induced by the electric field and is $\propto \tau$ because disorder is needed to keep the Fermi surface near equilibrium. Hence, the next term in the expansion will be formally of order zero in the disorder strength. This contribution is contained in vertex corrections in the diagrammatic formalism and the \textit{anomalous driving term} in the density matrix formalism of Refs.~\cite{Culcer2017, BurgosPRR2022}. In systems with spin or pseudospin degrees of freedom, such as those considered here, these corrections represent spin/pseudospin-dependent scattering phenomena commonly termed skew scattering and side jump \cite{Ado2015, Ado2016}. They are associated with the Fermi surface rather than the Fermi sea and occur because the scattering potential also contributes to band mixing and inter-band coherence. An alternative understanding is that disorder mixes the bands in the crystal, giving a correction to the wave function, and this results in a correction to the band-expectation values of certain physical observables, which is proportional to the disorder strength. At the same time, an electric field shifts the Fermi surface away from equilibrium, and this shift is inversely proportional to the disorder strength. The net result in the average of physical observables is the product of the correction to the band-expectation value and the shift in the distribution function, and this product is formally independent of the disorder strength.

Terms to zeroth order in the disorder strength play crucial roles in Hall effects, as has been shown for the spin Hall effect \cite{Inoue2004, Khaetskii2006, Gradhand2010, FerreiraAires2014, SinovaJairo2015, CullenJames2023}, as well as in the anomalous Hall effect \cite{Inoue2006, Sinitsyn_2007, BurgosPRR2022} and its non-linear counterpart \cite{SNandy2019, DuZZ2019, DuZZ2021NatureComm, OrtixCarmine2021, DuZZ2021, Burgos2023}. The only situation where disorder is irrelevant is the insulating state because disorder corrections are Fermi surface effects and vanish in the gap so that the intrinsic contributions dominate the response. In light of this and the fact that the OHE itself is an inter-band coherence phenomenon, disorder terms are expected to be important in the OHE in doped systems. Disorder has already been shown to play an important part in the orbital magneto-electric effect \cite{RouAndDimaPesin2017}.

There is an \textit{atomic} contribution to the OHE, and the disorder affects this contribution. In a recent paper, Pezo \textit{et al.} considered Anderson-type disorder in the atomic and itinerant contribution to the OHE \cite{PezoArmando2023}. The authors concluded that for various two-dimensional materials, the atomic contribution is more robust against disorder as compared to the itinerant contribution. Specifically, they considered graphene on h-BN and hydrogenated graphene. In both cases, $p_x-p_y$ hybridization is induced, and the itinerant and atomic effects coexist, and more importantly, SHE is also present. They showed that in both systems, the atomic response dominates. One more conclusion in Ref.~\cite{PezoArmando2023} is that the energy dispersion has a more dramatic effect on the orbital current in the presence of disorder as compared to its effect on topology. Such a conclusion is closely related to Ref.~ \cite{liu2023} since they concluded that disorder dominates over the intrinsic (topological) effect in the metallic regime. This can result from a diffusive regime, where the dispersion relation plays an important role. Moreover, Pezo et al. \cite{PezoArmando2023} also found that the size of the gap is crucial in the OHE, which also appears in Ref.~\cite{liu2023} as a signature of the inter-band coherence effect. In fact, from the anomalous Hall effect, it is expected that inter-band coherence mechanisms are present in the full
$\alpha^{(0)}_{ij}$ susceptibility. 

Vertex corrections embody the main effect of disorder on Hall conductivities. Although vertex corrections to the OHE in Si are known to vanish \cite{Bernevig2005}, they are generally present in the OHE in other materials. In Ref.~\cite{PezoArmando2023}, the disorder is introduced as a random function with a different value at each atomic site. While this is an excellent development, it does not account for vertex corrections, which are challenging and expensive to account for in computational approaches. A recent paper addressed the role of disorder in the OHE \cite{tang2024}, including the vital vertex corrections, applying it to a generic two-band $p$-orbital tight-binding model. The authors showed that at very specific symmetry points, disorder corrections can even cancel the intrinsic response. This situation is in marked contrast to disorder corrections to the OHE in massive Dirac fermion systems, where it was recently shown that disorder effects indeed dominate the response \cite{liu2023} and overwhelm the intrinsic contribution by a factor of $\approx 20$ for short-range scalar impurities. The channel $\alpha^{(-1)}_{ij}$ is expected to dominate as the ballistic regime is approached.

\subsection{Conservation of OAM}
\label{Sec:conservationOAM}

Three factors make OAM conservation crucial. Firstly, for information processing, it is imperative to know how long the OAM can live and which mechanisms cause it to change. In principle, spin and OAM dynamics can occur on different time scales. Consider the example of OAM injection into an FM and exerting an orbital torque on the local magnetisation. A theoretical prediction in Ref.~\cite{GoDongwook2023} shows that both the spin and the OAM affect the local magnetisation in a ferromagnet. However, increasing the thickness of the ferromagnet shows that the OAM effect persists longer than the spin effect, indicating that the OAM possesses a longer diffusion length. In this particular case the mechanism of orbital dephasing was attributed to hybridisation due to the crystal structure of the ferromagnet. 

Secondly, since OAM transport underpins the OHE and orbital torque, it is essential to know if the quantity transported is conserved. Research on the spin-Hall effect in spin-orbit coupled systems, in which the spin is not conserved, has revealed that defining a conserved spin current is exceedingly difficult \cite{ShuichiMurakami2003, ShiJunren2006, CullenJames2023, HongLiu2023-II}. As a result, considerable confusion remains in the field concerning the mechanisms dominating charge-to-spin conversion in different materials. Similar issues may arise in the OHE case if the relevant OAM is not conserved.

Finally, the orbital torque inherently requires OAM non-conservation since the OAM injected into a ferromagnet needs to be converted into spin angular momentum before it exerts torque on the magnetisation. This process requires spin-orbit coupling, and the interplay between the OAM, the electron spin, the magnetisation, and the details of the interface is highly non-trivial.

OAM conservation in centrosymmetric systems has been studied by Han \textit{et al.} in Ref.~\cite{SeungynHan2022} in equilibrium and in an electric field using a quantum Boltzmann formalism and accounting for the disorder. The authors argue that many aspects of OAM and spin-1/2 dynamics are qualitatively different and identify orbital responses with no spin counterpart. The primary reason is that angular momentum operators are incomplete, and products of operators are required to span the entire Hilbert space. The physics is strongly reminiscent of spin-3/2 systems such as holes in diamond and zincblende materials, which possess a quadrupole degree of freedom that makes spin dynamics highly non-trivial \cite{RolandWinklerBook}. The orbital torsion identified in Ref.~\cite{SeungynHan2022} can be regarded as a counterpart of this quadrupole degree of freedom. Interestingly, Han \textit{et al.} identify oscillations in the OAM expectation value that do not require ${\mathcal T}$ or ${\mathcal P}$-breaking, which is analogous to the alternating spin polarisation experienced by low-dimensional holes driven around a corner \cite{CulcerLechnerWinkler2006} and reflect the non-conservation of the OAM Bloch vector and OAM transfer between different multipoles. In this context, Ref.~\cite{SeungynHan2022} identified an orbital angular position operator (OAP). This element contains an even power of the OAM and has the form $H=\sum_{\alpha,\beta}h_{\alpha\beta}(\bm k)\{L_{\alpha}, L_{\beta}\}$, where the orbital texture $h_{\alpha\beta}(\bm k)$ is an even function of momentum. This contribution is expected to be relevant for orbital torque generation but remains to be experimentally determined.

OAM conservation has also been studied by Dongwook Go et al. in Ref.~\cite{GoDongwook2020-II} in the presence of spin-orbit coupling. The authors developed a continuity equation that includes sources of angular momentum transfer between different degrees of freedom in a system. Such a theory was developed to treat spin and orbital degrees of freedom on an equal footing. 
To this end, it is important to distinguish the angular momentum (spin and orbital) carried by Bloch electrons, the mechanical angular momentum of the lattice, and the spin angular momentum of the local magnetic moment of an FM. The theory then describes spin-angular momentum 
transferred to the local magnetisation through exchange interaction, the orbital angular momentum of the electron to the crystal lattice through 
crystal field potential, and the conversion of spin-angular momentum to orbital angular momentum through spin-orbit coupling. Fundamentally, the torque on the 
OAM occurs due to the crystal field and spin-orbit interaction, the last one being responsible for the generation of a spin-polarized current from an orbital-polarized one. The authors alsoaddressed the role of disorder in an attempt to satisfy the continuity equation for both inter-band and intra-band effects. They found that intrinsic processes related to inter-band transitions in the torque are balanced by extrinsic relaxation processes related to intra-band transitions. However, as the authors point out, their approach is phenomenological and does not consider side jump or skew scattering \cite{GoDongwook2020-II}.

In the context of valley physics, Kazantsev et al. \cite{kazantsev2023} showed that the valley density (related to OAM density \cite{BhowalSayantika2021}) in fully gapped time-reversal symmetric nanoribbons of graphene is not conserved out of equilibrium. 
They derived a continuity equation and chose a specific geometry for the edges so that the valley number is conserved in equilibrium. The system supports a valley Hall effect in the bulk driven by an external electric field. However, this transversal flow of \textit{valley density} does not necessarily lead to an accumulation of \textit{valley density} at the edges, which is somehow a paradox since the valley density must go somewhere. They solved the paradox by including the external electric field in their modified continuity equation, specified by
\begin{equation}
\label{Eq:Vignalecontinuityequation}
\frac{\partial n_{v} (y,t) }{\partial t}+\frac{\partial j^{y}_{v}(y,t)}{\partial y}=-e^2E_x(t)\sum_{\bm k} S(k)\frac{\partial f_{k}(y)}{\partial k},    
\end{equation}
where $n_{v}$ is the valley density, $j^{y}_{v}$ is the valley current density,  $S(k)=\pm 1$ is the valley charge, $f_{k}(y)$ is the electronic distribution, $e$ is the electron charge and $E(t)$ is the driving field. The authors showed that at zero temperature and in a fully gapped $\mathcal{T}$-symmetric insulator, the right-hand side cancels the second term on the left, and then there is no valley density accumulation. In the metallic regime, the system allows valley density accumulation, but it depends on dissipation, and more importantly, such an accumulation depends not only on the bulk valley Hall current generated by the electric field but also on the form of the wave functions at the boundaries. The authors concluded then that if a system exhibits a valley Hall effect it can not be a true insulator.

It is also important to study what are the mechanisms or symmetries that induce torque on the OAM of Bloch electrons in a specific system that can host an orbital current. To see this, consider the theory developed for OAM conservation in Ref.~~\cite{GoDongwook2020-II}, which holds a number of general lessons. The authors consider a general situation where angular momentum can be transferred from the electrons to the local moment of an adjacent FM or to the crystal lattice. There is an issue related to the diffusion of orbital polarized current in a NM. For instance, if an orbital current is generated by orbital pumping in an FM, such an orbital current can be transformed into a charge current, but first, it needs to travel through a material where the inverse OHE takes place \cite{Santos2023}. Ref.~\cite{Burgos2023nonconservation} showed that the OAM is, in general, not conserved out of equilibrium. At each $\bm k$, the torque on the OAM reads
\begin{widetext}
\begin{align}
\label{Eq:generatorque}
\tau^i_{nn;{\bm k}}
&=
\displaystyle \frac{e E_a}{\hbar }\epsilon_{ijk} \,\bigg( \sum_{m \ne n} 
{\rm Im}[\mathcal{Q}^{nm}_{ka} ] \, (v^j_{mm} + v^j_{nn}) - \sum_{m \ne n \ne l}  
{\rm Re}\bigg[
\frac{\mathcal{R}^{a}_{nl} \left(v^k_{lm} v^j_{mn} + v^j_{lm} v^k_{mn} \right)}{\varepsilon_n - \varepsilon_l}
\bigg] \bigg)
\end{align}
\end{widetext}
The above equation is general and can be applied to any system described by Bloch states. It represents the rate of change of the operator $\bm L=\bm r\times \bm v$ which is the fundamental quantity transported in the OHE, and which semi-classically is interpreted as the self-rotation of the wave-packet about its centre of mass. The first sum survives in a two-band model, whereas the second sum vanishes if the model spans only two bands. In the first sum, the factor in parenthesis is the sum of two group velocities in different bands. It is easily seen  that in two-band systems with particle-hole symmetry the torque is zero. The other terms represent the anomalous velocity that gives rise to the anomalous Hall current \cite{XiaoDi2010}. Then, the orbital moment is not conserved in $\mathcal{P}\mathcal{T}$-broken states while it is conserved in $\mathcal{P}\mathcal{T}$-symmetric. It should be noted that Eq.~\eqref{Eq:generatorque} can be interpreted as the effect of the electric field on the band structure of the system in the same way as the anomalous velocity. In this case, the main mechanism of \textit{randomization} is the coupling of neighboring bands by the external electric field. The effect is related to inter-band virtual transitions that cause a precession of the orbital moment of the order of nanoseconds. This effect is a dissipationless Fermi sea effect.

\section{Experimental observation} 
\label{Sec:Experiments}

\begin{figure}[tbp]
\centering
\includegraphics[width=1.05\linewidth]{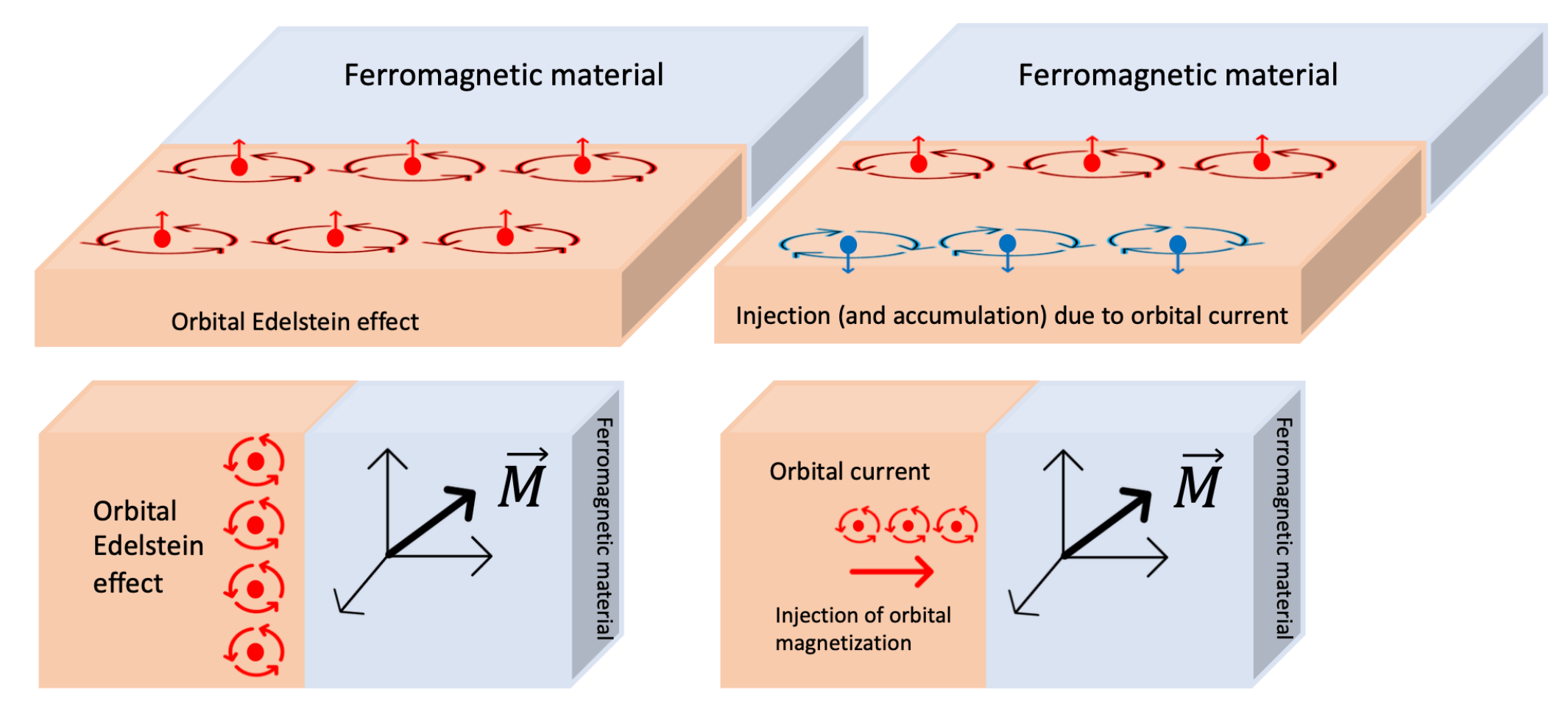}
\caption{Sketch of the orbital magneto-electric effect and orbital Hall effect. The orbital degree of freedom can interact with the local
magnetisation in a ferromagnetic material through the spin-orbit coupling in the ferromagnetic material or in the interface between them. It is essential to have orbital-to-spin conversion in order to induce an effect in the magnetisation of the FM.}
\label{Fig:applicationOMEandOHE}
\end{figure}

Non-equilibrium OAM densities have attracted considerable experimental attention with a view to applications. In this case, a \textit{charge-to-orbital} current conversion is needed via OHE or OME (see Fig.~\eqref{Fig:applicationOMEandOHE}). However, for applications such as electric manipulation of magnetisation, \textit{orbital-to-spin} current conversion is necessary because the orbital degree of freedom does not interact directly with the magnetisation of a ferromagnet. Most of the experimental work on the OAM measures its effect on an adjacent ferromagnet -- an \textit{indirect} interaction with the local magnetisation. This has been done, for instance, by injection of OAM from a light metal (i.e., with weak spin-orbit coupling) and orbital torque generation. An unexpected contribution of current-induced spin-orbit torque efficiency in materials with weak spin-orbit coupling is a signature of a mechanism beyond spin-related interaction. For example, this has been reported in Ref.~\cite{LeeSoogil2021, LeeDongjoon2021} and explained by orbital-related magnetism. On the other hand, a \textit{direct} measurement of orbital dynamics involves detecting the OAM directly, which can be accomplished through, e.g., Kerr rotation. The OME was detected directly in strained MoS$_2$ in Ref.~\cite{LeeJieun2017} via Kerr microscopy. Strain was used to lower the symmetry, allowing the magneto-electric effect to occur. The authors report the observation of \textit{valley magneto-electricity}, which is exactly analogous to the OME.

\begin{figure}[tbp]
\centering
\includegraphics[width=.8\linewidth]{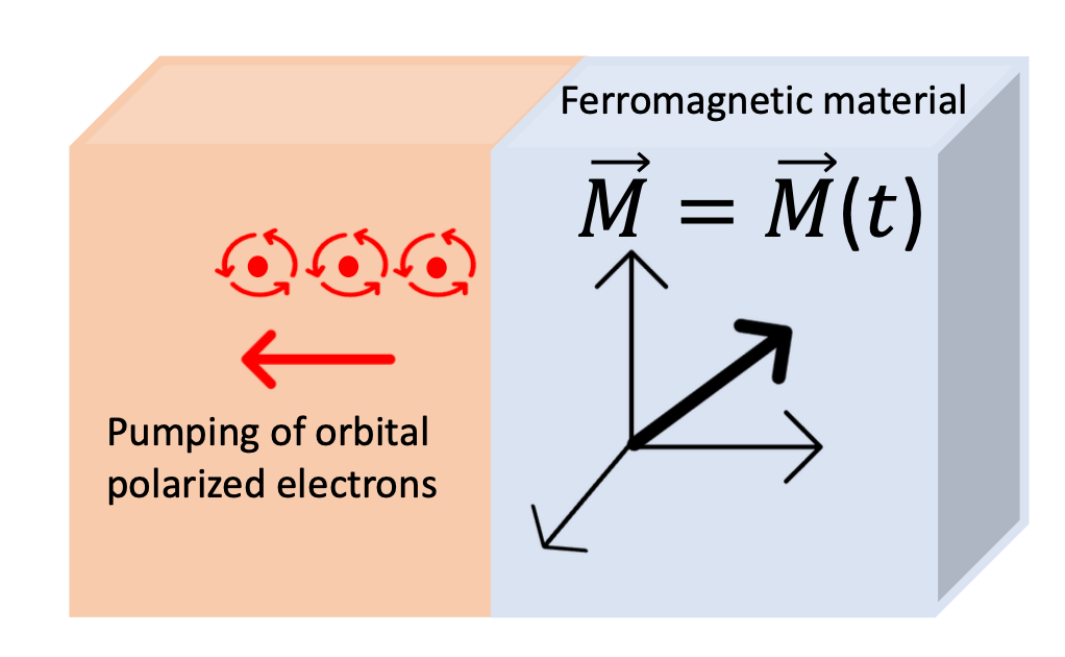}
\caption{Sketch of the orbital pumping effect. In the presence of dynamical magnetisation in an FM, orbital polarized current can be pumped from the FM to the nonmagnetic material.}
\label{Fig:pumpingeffect}
\end{figure}

Onsager reciprocal relations ensure that the inverse \textit{spin-to-charge} current conversion is possible, and for spin, the mechanisms are the inverse spin Hall effect (ISHE) and the inverse Edelstein effect (IEE). The orbital counterpart will be a conversion like \textit{spin-to-orbital-to-charge} current conversion. 
The inverse process of injection is then possible, whereby magnetisation dynamics \textit{pumps} an orbitally polarised current -- this is the inverse of the orbital torque and can be called an orbital pump. The basic idea is to excite the magnetisation in an FM to generate orbital and spin angular momentum 
(see Fig.~\eqref{Fig:pumpingeffect}). The FM is in contact with a NM material into which the orbital and spin angular momentum are pumped. After that, an electric current can be generated by the inverse orbital magneto-electric effect (IOME), the inverse orbital Hall effect (IOHE), or their spin counterparts. If the non-magnetic material has weak spin-obit coupling, the charge current is presumed to originate in the orbital degree of freedom. For instance, an orbital-to-charge conversion via IOME has been recently observed at the interface between LaAlO$_3$/SrTiO$_3$~\cite{ElHamdi2023}. This opens the possibility of IOME applications in 2DEGs \cite{Manipatruni2019}.

\subsection{Orbital torque generation by the OME}

Torque generation due to charge-to-spin conversion was reported in a ferromagnetic-metal/Cu/Al$_2$O$_3$ trilayer without heavy metals, whose origin is consistent with the orbital Rashba Edelstein effect \cite{KimJunyeon2021}. In Ref.~\cite{KimJunyeon2021}, the authors showed conversion of charge to orbital current via OME followed by orbital current transported and injected into an FM followed by orbital-to-spin conversion (due to spin-orbit coupling in the FM) and finally spin-orbit torque (electric manipulation of magnetisation) generation. Such a process is also called orbital torque (OT) since its starting point is charge-to-orbital current conversion rather than charge-to-spin current conversion. 
The authors of Ref.~\cite{ChenXi2018} reported large torque efficiency in the system Pt/Co/SiO2/Pt. The breaking of inversion symmetry and the 
hybridisation in Pt/Co and Co/SiO$_2$ produce the OME that transfers torque to Co. Shilei Ding et al. \cite{DingShilei2020} reported an enhancement of spin-orbit torque in TmIG/Pt by capping with a CuO$_x$ layer, namely in TmIG/Pt/CuO$_x$. This is consistent with a generation of an orbital polarised current in the 
interface of Pt/CuO$_x$ that is transformed to spin current in Pt and transported through it until it reaches TmIG to exert torque. The weak 
SOC of the interface allows the authors to infer the orbital origin of the effect.

Twisted bilayer graphene allows spontaneous orbital magnetization as indicated by measurements of a quantum anomalous Hall effect at 3/4 filling \cite{SharpeAaron2019, Serlin2020}. The orbital origin of the effect is inferred due to the weak SOC in the systems. The importance of such a system stems from the fact that a small current can switch the magnetisation to the opposite direction, then allowing an electric control of magnetisation. Ref.~\cite{HeWenYu2020-II} showed that the reduction of the lattice symmetry by the twist and the hBN substrate allows out-of-plane orbital magnetisation. This is possible given the large Berry curvature of flat bands at magic angles induced by the twist (we recall that one contribution to the OAM is closely related to the Berry curvature). The substrate also plays a fundamental role since it is responsible for inducing the gap and strain.

\subsection{Orbital torque generation by the OHE}

Injection of an orbital current into an FM using the OHE is expected to generate an OT on the local magnetic moments. The OHE causes electrons with opposite OAM to accumulate on opposite sides of a sample, then creating an OAM  density at the edges. If this sample is in contact with a magnetised material, the OAM will exert an orbital torque on the magnetisation. 
The effect, in principle, does not require SOC because the electric field couples directly to the orbital degree of freedom, in contrast to the spin degree of freedom where the electric field couples to the spin through SOC. Then, materials with weak SOC are a source for OHE realization. Zr is a light element with weak spin-orbit coupling. The authors of Ref.~\cite{Zheng2020} showed magnetisation switching is possible in multilayer Zr/CoFeB/MgO, where the OHE is induced in Zr. Cr is likewise a light element that can be used to generate OHE. This case was achieved in Ref.~\cite{LeeSoogil2021}. The authors placed heavy elements like Gd or Pt between Cr and the FM so that an orbital-to-spin current conversion happens before it is injected into an FM to produce OT. Bose et al. in Ref.~\cite{BoseArnab2023} used bilayers of (Nb or Ru)/Ni and (Nb or Ru)/FeCoB and reported torque on the magnetisation in the FM where the orbital to spin conversion happens in the FM. They also considered the case of Pt between the NM and the FM. This is to achieve
orbital to spin conversion in Pt. Similarly, Ref.~\cite{LiuFu2023} also considered OT generation in bilayers of Nb/ (Co or Fe), where the orbital-to-spin current conversion occurs in the FM.

It is worth noting that the OHE also occurs in NM materials with SOC, as shown in Ref.~\cite{LeeDongjoon2021}. Actually, in such a case, OHE and SHE compete, and the dominant channel for torque generation in an adjacent FM is determined by the spin-orbit correlation in the NM and the FM \cite{LeeDongjoon2021}. This happens because orbital current can be directly injected into the FM, and due to the spin-obit correlation in the FM, the orbital 
current becomes spin current that finally exerts torque.
A recent paper \cite{GiacomoSala2022} identified a giant orbital Hall current in Cr and a strong orbital Hall current in Pt, which was considered as a source of only spin Hall current. The experiment shows that the spin-orbital diffusion length is longer in light metal (Cr) as compared to heavy metal (Pt), then the efficiency of the torque generation in an adjacent FM can be optimized by increasing the thickness of the NM material, and this can be even larger than in Co/Pt bilayers.

Knowing for how long an orbital polarized current can travel is then a subject of paramount importance for orbitronic applications.
Long-range orbital transport has also been experimentally inferred in Ref.~\cite{DuttaSutapa2022, LiaoLiyang2022, BoseArnab2023, LiuFu2023}.
In particular Ref.~\cite{LiaoLiyang2022} reports torque generation in CoFeB/Ru/Al$_2$O$_3$: orbital current is generated in Ru/Al$_2$O$_3$ and transported to 
CoFeB. The torque increases with increasing the thickness of CoFeB. The effect of FM thickness on OT generation was also considered in Ref.~\cite{BoseArnab2023} for the case of NM/Ni systems.

In Ref.~\cite{HayashiHiroki2023}, the authors show that orbital current can propagate over long distances in non-magnetic (NM) and ferromagnetic materials (FM).
The authors used a bi-layer of FM/NM material. In particular, they used the light metal Ti for NM, where the spin Hall current is believed to be 
smaller than the orbital Hall current \cite{DuChunhui2014} (and also opposite in sign). For the FM, they used Ni or Ni doped with Fe, where the doping effect is to lower the efficiency of orbital-to-spin conversion since it affects the FM's electronic structure. Their results agree with spin-torque ferromagnetic resonance predictions \cite{FangD2011, LiuLuqiao2011}. Specifically, the authors of Ref.~\cite{HayashiHiroki2023} found a damping-like torque whose sign is consistent with OHE in the Ti layer rather than with predictions for SHE. Moreover, they also found that the order of magnitude of the 
damping-like torque is at least two orders of magnitude higher than the predicted due to SHE in Ti. These experiments provide indirect evidence of the dominance of the orbital degree of freedom in the bi-layer of FM/NM. Recently, however, the OHE was measured by directly detecting the OAM accumulation at the edges using Kerr rotation in~\cite{IgorLyalin2023} for Cr and in~\cite{ChoiYoungGwan2023} for Ti.

An important issue from the point of view of applications is the efficiency of the process of converting the orbital degree of freedom in order to exert a torque on an adjacent FM. In table \eqref{tab:orbitalconversion} we present experimental values reported in OHE studies. In particular the first line reported values for Pt/TmIG, where only (or mainly) the spin degree of freedom is believed to be important. A clear enhancement is shown in Ta(8)/Pt/TmIG when an $8$nm thick layer of Ta is placed to form a trilayer. Such an enhancement is attributed to the generation of OAM in Ta.

\begin{table}[tbph]
\begin{ruledtabular}
\begin{tabular}{ l c c c  }
Material & $\xi_{DL}$ & $\xi^{E}_{DL} (\Omega^{-1}m^{-1})$ \\
\colrule
Pt/TmIG \cite{Tianhui2023} & 0.025 & $0.326\times 10^{5}$ \\
Ta(8)/Pt/TmIG \cite{Tianhui2023} & 0.46 &  $2.26\times 10^{5}$ \\
Nb/FeCoB \cite{BoseArnab2023}  & $-0.001\pm0.001$ & $(-2.4\pm 2.4)\times 10^{3}$ \\
Ru/FeCoB \cite{BoseArnab2023}  & $0.001\pm0.001$ & $(3.6\pm 3.6)\times 10^{3}$ \\
Nb/Ni  \cite{BoseArnab2023}    & $0.018\pm 0.005$ & $(44\pm 12)\times 10^{3}$ \\
Ru/Ni \cite{BoseArnab2023}  & $0.019\pm 0.005$ & $(69\pm 18.3)\times 10^{3}$  \\
Ru/Pt/FeCoB/Pt \cite{BoseArnab2023}  & $ 0.03\pm 0.004$ & $(110\pm 15) \times 10^{3}$ \\
\end{tabular}
\end{ruledtabular}
\caption{\label{tab:orbitalconversion} Torque generation efficiency $\xi_{DL}$ and torque generation efficiency per unit electric field $\xi^{E}_{DL}$ for different materials 
as reported in experiments. In Ref.~\cite{BoseArnab2023} the values are reported in 
units of $(\hbar/2e)$. Also, in Ref.~\cite{BoseArnab2023} the effect of placing Pt on both sides of FeCoB is to cancel the effect of the spin Hall contribution.
The importance of the orbital degree of freedom is 
explict in this table as reported in Ref.~\cite{Tianhui2023} in the first materials: once an $8$nm thick Ta is used, the efficiencies reported are enhanced. The role of the 
element Ta is to generate orbital Hall effect that will be injected into the FM and eventually will exert an orbital torque. In principle, without Ta, the torque is solely due to the injection of spin Hall current.} 
\end{table}

\subsection{Pumping of orbitally polarized charge current and IOME}

The generation of a charge current from a spin-polarized current can be achieved by the Edelstein effect as shown in Ref.~\cite{Sanchez2013} and is called the inverse Rasbha Edelstein effect. A similar phenomenon is expected in the context of orbital magnetisation. The inverse OME, or the OAM-to-charge conversion, was observed recently in Ref.~\cite{ElHamdi2023}. The authors generated orbital and spin-polarized currents in a ferromagnet by spin pumping and the spin Seebeck effect, where a temperature gradient drives spins. The current is injected into the 2DEG in LaAlO$_3$/SrTiO$_3$, and a voltage is measured. The effect was predicted theoretically in Ref.~\cite{JohanssonAnnika2021}.

\begin{figure}[tbp]
\centering
\includegraphics[width=\linewidth]{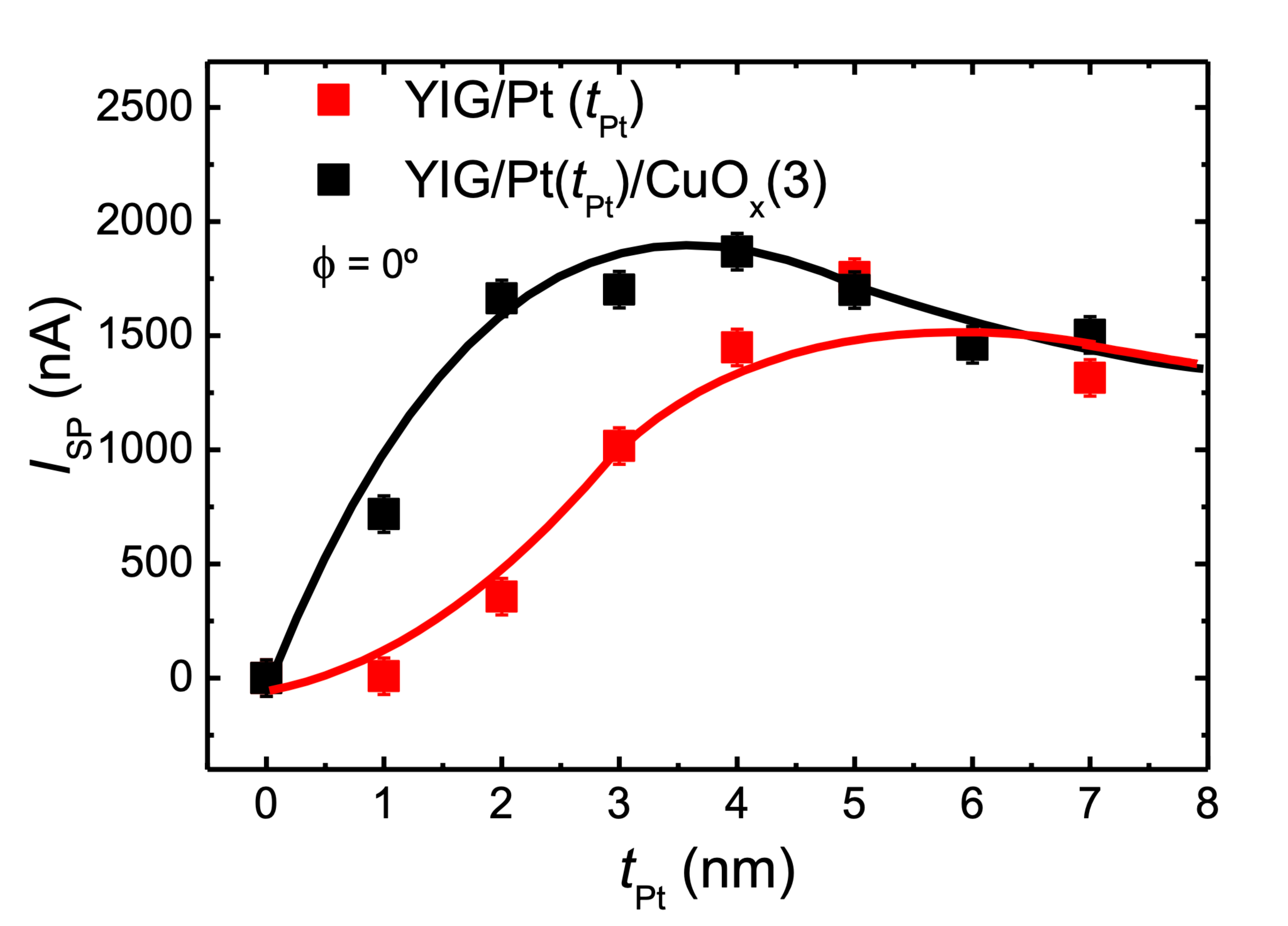}
\caption{The variation of the spin current signal with the thickness of Pt, as reported in Ref.~\cite{Santos2023}. It shows a clear difference between 
Y$_3$Fe$_5$O$_{12}$/Pt/CuO$_{\rm x}$ and  Y$_3$Fe$_5$O$_{12}$/Pt. Including CuO$_{\rm x}$ adds a contribution to
orbital-to-charge conversion arising from OEE. This depends on the thickness of the material because the orbitally polarised signal must 
travel through Pt until it reaches the oxidized interface. In  Y$_3$Fe$_5$O$_{12}$/Pt/CuO$_{\rm x}$ the signal start to decrease after 
$t_{Pt}\sim 3nm $. Beyond this thickness, both elements behave the same because the orbital current is dephased before getting into 
the Pt/CuO$_{\rm x}$ interface. A similar but more general study was performed in Ref.~\cite{Santos2024,santos2024arxiv}.}
\label{Fig:acevedopaper}
\end{figure}

One more experiment that measures the inverse orbital torque and can be considered as the orbital counterpart of the inverse Rasbha Edelstein effect was considered in Ref. \cite{Santos2023}. They used a trilayer of Y$_3$Fe$_5$O$_{12}$/Pt/CuO$_{\rm x}$ and a bilayer of Y$_3$Fe$_5$O$_{12}$/Pt. In the trilayer, CuO$_{\rm x}$ stands for surface-oxidized Cu, and the inversion symmetry breaking at the interface Pt/CuO$_{\rm x}$ is known for generating the orbital Rashba Edelstein effect \cite{DingShilei2020, GoDongwook2021, FukunagaRiko2023}. The way the authors in Ref. \cite{Santos2023} showed the orbital origin of the inverse torque effect is as follows: ferromagnetic resonance is used to pump spin from the ferromagnet Y$_3$Fe$_5$O$_{12}$ into Pt. Due to the strong spin-orbit coupling in Pt, the ISHE takes place, and a charge current can be measured. Then, an extra layer of CuO$_{\rm x}$ is added to the system, and an increased current is measured this time. This extra signal occurs because of an IOHE induced by the orbital Rashba Edelstein effect at Pt/CuO$_{\rm x}$. The measured voltage decreases in the trilayer Y$_3$Fe$_5$O$_{12}$/Pt/CuO$_{\rm x}$ when the thickness of Pt is increased. This shows that the orbital polarized current must reach the interface Pt/CuO$_{\rm x}$ where the orbital-to-charge current conversion happens, meaning that the diffusion length of the orbital current plays an important role (See Fig.\eqref{Fig:acevedopaper}).

\subsection{Pumping of orbitally polarized charge current and IOHE}

\begin{figure}[tbp]
\centering
\includegraphics[height=10.5cm]{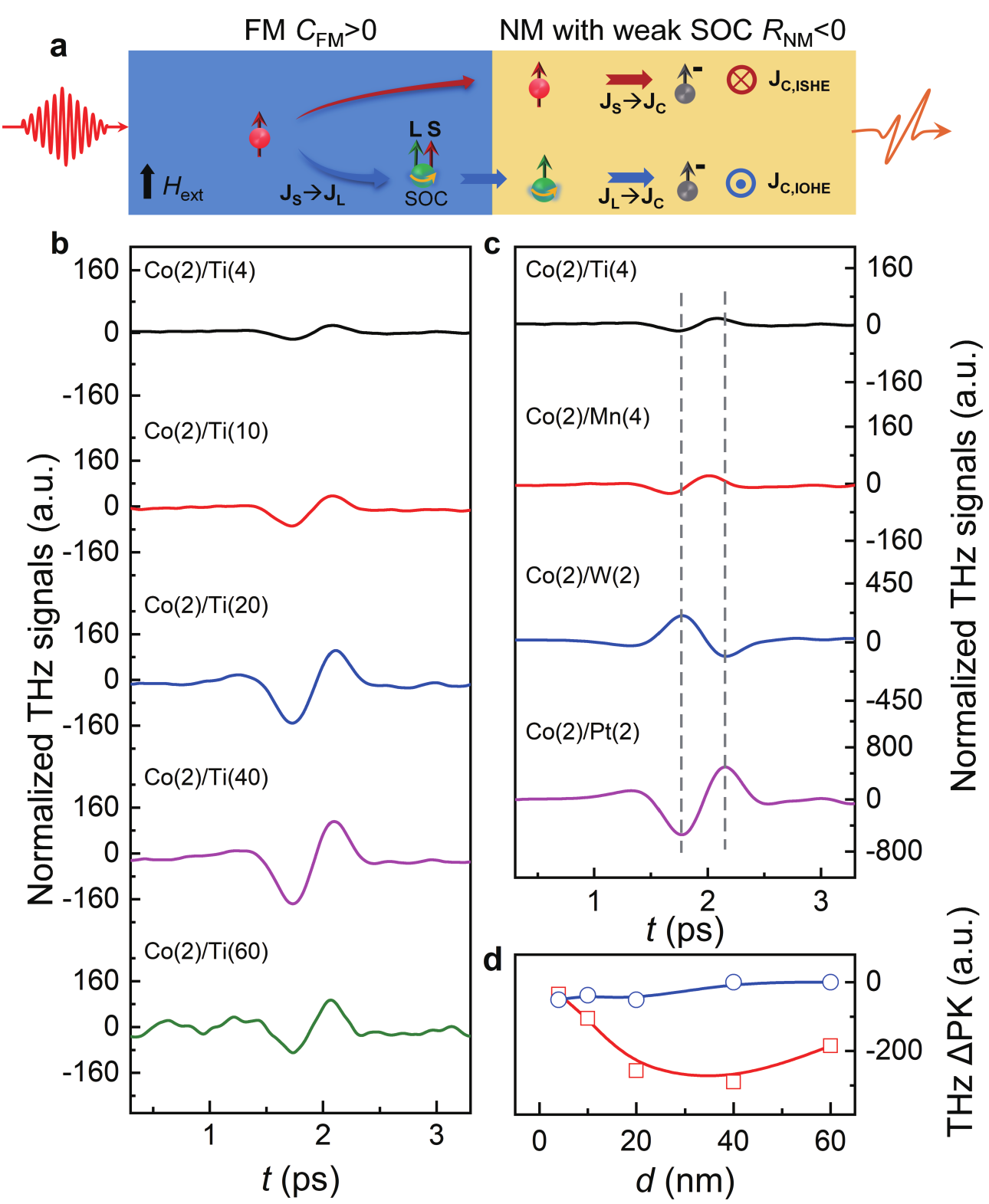}
\caption{ a) Spin-to-charge and orbital-to-charge current conversion mechanisms in ferromagnetic (FM) and non-magnetic (NM) materials. b)-c) THz signal is applied to the FM
to produce spin and orbital pumping in Ref.~\cite{WangPing2023}. 
The signal shows inverse orbital Hall effect (IOHE) in Co/Ti and Co/Mn, along with inverse spin Hall effect (ISHE) in Co/W and Co/Pt. 
The number in parenthesis beside the elements indicates their thickness in nm. 
d) The signal shows that the diffusion length in Co/Ti is longer than in Co/Mn, indicating that the process of orbital-to-charge conversion is more efficient in Co/Ti. This figure is adapted from Ref.~\cite{WangPing2023}. }
\label{Fig:PingWangPaper}
\end{figure}

A recent experiment on the bi-layer of Ti/Co and Mn/Co  \cite{WangPing2023} reports evidence of IOHE. In the experiment, a femtosecond laser is pumped into the Co layer to produce a spin current via the exchange interaction. Spin-orbit coupling in the Co leads to spin-to-orbital current conversion. Then, both 
spin and OAM are pumped into the NM layer, where both ISHE and IOHE take place, and the signal is measured. Given the small spin-orbit coupling in Ti and Mn, the signal is dominated by the IOHE, although ISHE is also present, with the IOHE and ISHE signals predicted to be opposite in sign. Then, the authors used this property further to investigate the orbital origin of the terahertz signal as follows: the signal in a different bilayer of Pt/Co is dominated by the ISHE because of the strong SOC in Pt and is predicted to be opposite to the (small) signal due to ISHE in Ti/Co and Mn/Co. However, the experiment shows that the three signals Ti/Co, Mn/Co, and Pt/Co have the same polarity, meaning that the mechanism in the light metals Ti and Mn must be distinct from SHE and is, in fact, consistent with orbital mechanisms. The authors also studied the behavior as a function of Ti thickness and found agreement with orbital mechanisms reported in Ref.~\cite{ChoiYoungGwan2023} about orbital diffusion (See Fig.~\ref{Fig:PingWangPaper}). 
Seifert et. al. also showed the importance of the orbital degree of freedom in the generation of orbital polarised current by pumping in Ref.~\cite{SeifertTom2023}. The authors showed that orbital current generated for instance in Ni can travel through W 
with dacayment lenght of the order of $80$nm. Such an orbital current can be 
converted into a charge current at the interface W/SiO$_2$ by means of the inverse OME.

\subsection{Magneto-resistance measurement due to OAM}

Spin magneto-resistance and Hanle magneto-resistance are well-known effects due to the spin degree of freedom \cite{Dyakonov2007, ChenYanTing2013, Nakayama2013, Velez2016, YanTingChen2016, Zhang2019}, 
both resulting in a modification of the edge spin accumulation by a magnetic field. If the non-magnetic material where the spin accumulation forms is in contact with a ferromagnet, the external magnetic field can rotate the local magnetisation in the FM.  
Depending on the relative angle between the polarization of the spin and the local magnetisation, the current can be absorbed (for minimum magneto-resistance) or reflected (for maximum magneto-resistance) and converted back to a charge current (the inverse spin Hall effect). The Hanle effect is similar in nature and happens without an FM, the external magnetic field interacting directly with the spin accumulation. This effect is expected to be relevant for orbital magnetisation as well.

Magneto resistance at interfaces due to orbital mechanisms has been observed recently. Ding \textit{et al} \cite{DingShilei2020} used a trilayer of (TmIG)/Pt/CuO$_x$. Due to inversion symmetry breaking 
in the interface Pt/CuO$_x$, a non-equilibrium orbital magnetisation density develops. The OAM is injected into Pt, and then the orbital-to-spin conversion process occurs due to the SOC in Pt. The angular momentum is transported through Pt until it reaches the FM, whereupon an enhancement of the spin-orbit torque was observed. 
Such measurements are corroborated by magneto-resistance by rotating the local magnetic moment in the 
FM with an external magnetic field. A similar effect was observed in a bilayer of Permalloy/CuO$_x$ \cite{DingShilei2022}. The orbital counterpart of the Hanle magneto-resistance was recently demonstrated in single layer Mn in Ref. \cite{GiacomoSala2023}, where the OAM accumulated due to OHE is induced to change due to an external magnetic field. The size of the effect in Mn is comparable to that in Pt and Ta, where the accumulation is due to SHE.

We conclude this section with table \eqref{tab:experimentalparameters} of parameters relevant to OAM and spin applications in spintronics. In principle, the orbital degree of freedom dominates in Ta. Experiments show that this channel can travel longer as compared to the spin, as reported for instance for Pt, where the spin degree of freedom is believed to dominate.

\begin{table}[tbph]
\begin{ruledtabular}
\begin{tabular}{ l c c c  }
Material & $\lambda$ (nm) & $D$(m$^2$/s) & $\theta_{H}$ \\
\colrule
Mn \cite{GiacomoSala2023} & 2 & $2.5\times 10^{-5}$ & 0.016  \\
Pt(7nm)\cite{Velez2016} & $0.9$ & $6\pm 1 \times 10^{-6}$ & 0.056  \\
Ta \cite{Velez2016} & $1.0\pm 0.1$ & $(2.1\times 10^{-6}$ & -($0.008 - 0.15$) \\
                    &               & $-3.7\times 10^{-5})$ &  \\
Ta \cite{Hahn2013} & $1.8\pm 0.7$ & - & -0.02
\end{tabular}
\end{ruledtabular}
\caption{\label{tab:experimentalparameters}
Characteristic diffusion length $\lambda$, 
diffusion coefficient $D$ and Hall angle $\theta_{H}$ measured in recent experiments.}
\end{table}

\section{Summary, challenges and outlook}
\label{Sec:summary}

The developments of the last five years have marked the arrival of orbitronics onto the global stage. The main focus of the field to date has been on metals, semimetals, and semiconductors, with most of the experimental effort concentrated on multi-layers that exhibit OHE, OME, or their reciprocal counterparts. In addition to these, we anticipate the study of orbitronic phenomena to grow in superconductors, Moir\'e materials, and nonlinear effects. Superconductors offer the possibility of OAM generation by a dissipationless supercurrent, while orbital torque generation in Moir\'e materials brings up the intriguing prospect of testing the interaction between two OAM densities -- an equilibrium OAM magnetisation density stabilised by interactions and a non-equilibrium OAM density generated by an electrical current. At the same time, nonlinear responses remain a very active research area \cite{Sodemann2015, SNandy2019, PhysRevLett.129.227401, Sinha22, Chakraborty_2022, Burgos2023, PhysRevB.107.165131, KamalDas2023-II, mandal2024quantum}, which has extended into spintronics and orbitronics. Theoretical works in nonlinear regimes already pointed out the importance of the OAM of Bloch electrons in magneto-resistivity and valley transport \cite{Lahiri2022, KamalDas2023}. 
In \cite{KamalDas2023}, the OAM was shown to have a correction due to the applied electric field, leading to a nonlinear valley Hall effect, while the non-linear magnetisation in centro-symmetric systems was recently considered in Refs.~\cite{XiaoCong2022, XiaoCong2023,BejaminFregoso2022, baek2024nonlinear}. A time-reversal symmetric non-linear Hall effect was recently discovered in in LaAlO$_3$/SrTiO$_3$, which was attributed to the orbital degree of freedom~\cite{Lesne2023}, specifically to an orbital Rashba-like coupling -- the same mechanism responsible for IOME \cite{ElHamdi2023}. The same effect has been shown to lead to orbital photocurrents \cite{Mokrousov_Nonlinear_PRL2024}. Interestingly, it was recently suggested that an orbital-driven Kerr effect in inversion-breaking non-magnetic materials can be used to probe the non-linear Hall effect \cite{Ovalle_OKE2023}. With the understanding of orbitronic nonlinear effects in its infancy, we expect this aspect of the field to grow in the near future.

At the same time, many fundamental physical issues in orbitronics require a deeper understanding. Experimentally distinguishing orbital and spin effects remains a significant and long-standing challenge since spin and orbital currents are generally produced together in the materials under study -- for example, the surface states of Bi$_2$Se$_3$ exhibit a strong OAM \cite{ParkSeungRyong2012}. This observation applies both to direct methods, such as Kerr rotation, and to indirect methods, such as measuring magnetisation dynamics. An additional challenge for both experiment and theory is that the microscopic mechanisms leading to charge-to-orbital-to-spin conversion in the vicinity of interfaces are far from being understood. Similarly, the relationship between the OHE and accumulated OAM density at the edge of a sample is poorly understood.

In spin-orbit coupled materials, the degree to which the OAM coexists with spin-orbit coupling remains to be established. In many inversion breaking systems the OAM may stem directly from the spin-orbit interaction. Thus it is imperative to understand the interplay of OAM and spin dynamics, particularly the possibility that OAM effects may exceed spin effects in spin-orbit coupled systems. In fact the relationship between the OAM and spin-orbit coupling in general has only received sporadic attention. For example, a chiral OAM state was identified on the surface of metals with a generic multi-orbital band structure and proposed as the origin of the Rashba effect present on these surfaces \cite{ParkSeungRyong2011}. The idea was extended to magnetic metals shortly thereafter \cite{Jin-HongPark2013}, where it was shown that spin polarised bands in magnetic metals are frequently associated with chirality in the OAM structure. At the same time, using angle-resolved photoemission (ARPES),  Ref.~\cite{ParkSeungRyong2012} reported the observation of OAM in the surface states of Bi$_2$Se$_3$. This observation, confirmed by numerical calculations, provides evidence that, since OAM and spin angular momentum are essentially inseparable in spin-orbit-coupled systems, spin-momentum locked states exhibit concurrent chirality in their OAM structure. This aspect will need to be investigated thoroughly in the future. 

Theoretically, the biggest challenge is handling the position operator, the quantum mechanical treatment of which faces two difficulties. One is that the position operator is ill-defined in periodic systems since Bloch electrons are delocalised \cite{RestaRaffaele1998, Vanderbilt2018, Resta2018}. In particular, its band-diagonal terms are especially problematic. Yet these band-diagonal terms lead to Fermi surface, group velocity, and dipolar effects and must be included in a complete treatment. This is related to the second difficulty, namely the choice of a frame of reference, which in the semi-classical approach is overcome by calculating the OAM with respect to the center of a wave packet as shown in Fig.~\eqref{Fig:wavepacketrotation}. Developing a quantum mechanical theory based on localised Wannier functions as done in \cite{Thonhauser2005} is possible. However, it is not possible to use Wannier states to describe all systems \cite{Thonhauser2006}. The modern theory of polarisability \cite{Vanderbilt1993, King-Smith1993, Vanderbilt2018} provided the first quantum mechanical approach for treating dipole-like operators and can be extended to the OAM, while the quantum-thermodynamical approach of Ref.~\cite{ShiJunren2007} avoids the use of the position operator. However, these theories were developed for clean equilibrium systems.

A complete quantum mechanical theory of non-equilibrium OAM dynamics, a \textit{modern theory of the non-equilibrium} OAM, is currently unavailable and will need to be constructed from the ground up. This process requires closing existing gaps in the equilibrium theory as well, namely developing an understanding of disorder and inhomogeneities and elucidating the relationship between the modern theory and the effective theory developed for centro-symmetric systems. A modern theory of the non-equilibrium OAM is needed to determine when the OAM is conserved in an arbitrary system, what mechanisms can exert a torque on it, and the interplay between Fermi surface and Fermi sea effects. Likewise, it is essential to determine how much of the OAM is topological and which quantum geometric quantities contribute the most -- for example, the quantum metric tensor and the Berry curvature dipole, what role is played by disorder in the steady state and by boundaries and interfaces in inhomogeneous systems, as well as the way the OAM is influenced by electron-electron interactions, with particular emphasis on orbital magnetism \cite{HankeJanPhillipp2017, Grytsiuk2020, Twistnoics-Allan}.

In summary, the past half-decade has witnessed impressive progress in orbitronics, with the experimental generation of steady-state OAM densities and currents proceeding at full throttle. The absence of a fundamental understanding of the mechanisms directly leading to the substantial OAM densities observed recently is not an obstacle in the development of new devices, and we expect orbitronic devices to supplement spintronic functionalities in the foreseeable future. On the theoretical front, the need for a deeper understanding of OAM conservation and orbital-to-charge conversion, as well as disorder and inhomogeneities, provides significant motivation and inspiration. One of the salient lessons from orbitronics is that in modern materials, the charge, spin, orbital, and valley degrees of freedom are all intertwined, and the dynamics of one are very rarely independent of the dynamics of the others. Orbitronics has emerged as one of the most active research areas in modern condensed matter physics and will continue to inspire the exploration of fundamental physics and  development of advanced materials and technologies.

\acknowledgments
The authors acknowledge stimulating discussions with Raffaele Resta, Daniel Arovas, Misha Titov, Yuriy Mokrousov, Aurelien Manchon, Shuichi Murakami, Hiroshi Kohno, Giovanni Vignale, Roberto Raimondi, Thierry Valet, Wang Yao, Kam Tuen Law, James Cullen, Hong Liu, and Cong Xiao. This work is supported by the Australian Research Council Centre of Excellence in Future Low-Energy Electronics Technologies, project number CE170100039.

\bibliography{reviewpaper}

\end{document}